\documentclass[fleqn,usenatbib]{mnras}

\usepackage[T1]{fontenc}
\usepackage{ae,aecompl}

\usepackage{graphicx,color}	
\usepackage[dvipsnames]{xcolor}
\usepackage{amsmath}	
\usepackage{amssymb}	
\usepackage{times}
\usepackage{threeparttable}
\usepackage{epstopdf}
\usepackage{url}
\usepackage{bm}
\usepackage{hyperref}

\usepackage{graphicx}	
\usepackage{amsmath}	
\usepackage{tabularx}
\usepackage{amssymb}	
\usepackage{ulem}		
\usepackage{enumitem}	
\usepackage{booktabs}	
\usepackage{float}
\usepackage[utf8]{inputenc}

\newcommand{\eqb}{\begin{eqnarray}}
\newcommand{\eqe}{\end{eqnarray}}
\newcommand{\sth}{\sigma_{\rmn T}}

\newcommand{\tcr}{t_{\rmn {cr}}}

\newcommand{\lB}{\ell_{\rmn B}}

\newcommand{\gpmax}{\gamma_{\rm p, \max}}

\newcommand{\lp}{\ell_{\rm p}} 

\title[A roadmap to hadronic supercriticalities]
{A roadmap to hadronic supercriticalities: a comprehensive study of the parameter space for high-energy astrophysical sources}

\author[Mastichiadis et al.]{Apostolos Mastichiadis$^{1}$\thanks{E-mail: amastich@phys.uoa.gr},
Ioulia Florou$^{1}$,
Elina Kefala$^{1}$\thanks{Current Address: Departament de Física Quàntica i Astrofísica, Institut de Ciències del Cosmos (ICCUB), Universitat de Barcelona (IEEC-UB), Martí i Franquès 1, 08028 Barcelona, Spain}, Stella S. Boula$^{1}$, \newauthor Maria Petropoulou$^{2}$\thanks{E-mail: m.petropoulou@astro.princeton.edu}
\\ \\
$^{1}$Department of Physics, National and Kapodistrian University of Athens, Panepistimiopolis, GR 15783 Zografos, Greece \\
$^{2}$Department of Astrophysical Sciences, Princeton University, Princeton, NJ 08544, USA \\
}

\date{Accepted XXX. Received YYY; in original form ZZZ}

\pubyear{2020}

\begin{document}
\label{firstpage}
\pagerange{\pageref{firstpage}--\pageref{lastpage}}
\maketitle
\begin{abstract}
Hadronic supercriticalities are radiative instabilities that appear when large amounts of energy are stored in relativistic protons. When the proton energy density exceeds some critical value, a runaway process is initiated resulting in the explosive transfer of the proton energy into electron-positron pairs and radiation. The runaway also leads to an increase of the radiative efficiency, namely the ratio of the photon luminosity to the injected proton luminosity.
We perform a comprehensive study of the parameter space by investigating the onset of hadronic supercriticalities for a wide range of source parameters (i.e., magnetic field strengths of $1$ G$-100$ kG and radii of $10^{11}$~cm$-10^{16}$ cm) and maximum proton Lorentz factors ($10^3-10^9$). We show that supercriticalities are possible for the whole range of source parameters related to compact astrophysical sources, like gamma-rays bursts and cores and jets of active galactic nuclei. We also provide an in-depth look at the physical mechanisms of hadronic supercriticalities and show that magnetized relativistic plasmas are excellent examples of non-linear dynamical systems in high-energy astrophysics.
\end{abstract}

\begin{keywords}
instabilities -- radiation mechanisms: non-thermal -- galaxies: active -- gamma-ray burst: general
\end{keywords}


\section{Introduction}\label{Sec:1}
The non-thermal extended photon spectra of high-energy emitting sources, such as active galactic nuclei (AGN) and gamma-ray bursts (GRBs), suggest the acceleration of radiating particles to ultra-relativistic energies. Although leptonic models are widely favoured in explaining the observed $\gamma-$ray flux through the emission of relativistic electron-positron pairs, hadronic scenarios that invoke relativistic protons (and heavier nuclei) remain a viable alternative, while having direct implications for the sites of cosmic-ray acceleration and neutrino production \citep[e.g.,][]{Mannheim93, Halzen97,Mucke01,Bottcher98, Kazanas02, Asano07, Murase08} . Especially after the discovery of the astrophysical high-energy neutrino flux by IceCube \citep{Ice13} and the most recent association of high-energy neutrinos with the $\gamma-$ray emitting blazar TXS 0506+056 \citep{IC18a,IC18b}, hadronic models became more relevant than ever before \citep{DeMu14,P15,Cerruti15, kei18,Cerruti18,2019XPO,2019Pms,oik19,Murase19,Kimura19}.

The basic premise of a hadronic scenario as applied to the compact high-energy emitting region(s) of an astrophysical source can be summarised as follows. The model assumes the presence of a relativistic proton population that interacts with its environment in two main ways. First, the gyromotion of protons in the magnetic fields of the source produces synchrotron emission, and secondly, the photohadronic interactions with low-energy photons (henceforth, soft photons) lead to the production of many secondary particles. More specifically, photopair (Bethe-Heitler) production creates relativistic electron-positron pairs, while photopion production\footnote{There is another category of hadronic models, where pions are produced via inelastic proton-proton collisions, which have been mostly applied to microquasars \citep{Romero3, Romero5,BRR5}.} injects charged and neutral pions ($\pi^\pm$, $\pi^0$). The pions decay almost instantaneously to create very energetic $\gamma-$rays from $\pi^{\rm 0}$ decay and more electron-positron pairs from $\pi^{\rm \pm}$ decays. $\gamma-$rays can, also, interact with soft photons via $\gamma \gamma$ absorption, thus producing more pairs. Neutrons and neutrinos are another by-product of photopion interactions. While neutrinos will escape from the source, neutrons may interact with ambient photons in a similar way as protons, decay inside the source or decay after they escape from the source.

These interactions, which relate all stable particle species to one another, form a non-linear network of physical processes. Their non-linear character is manifested more strongly when the proton energy losses are not driven by external agents, like magnetic fields and/or external radiation fields. In this case, feedback loops based on different combinations of the processes described above might become operative, if certain conditions are fulfilled. This may happen, for example, if the injection of pairs produced from photohadronic interactions significantly increases the number of soft photons in the source (e.g., by synchrotron radiation), which, in turn, serve as targets for further cooling of the protons. This feedback network has been first studied by \cite{Stern91} and \cite{Stern92} by means of Monte Carlo simulations, and later by \citet{KM92,MK05} and \citet{PM12} using the kinetic equation approach. Similarly, when the luminosity of the proton-injected $\gamma-$rays grows sufficiently, this high-energy radiation is automatically quenched, providing a supplementary soft target photon population for proton cooling, which results in more $\gamma-$rays \citep{Stawarz07, PM11, Petropoulou13}. 

As it was shown analytically in \cite{KM92} and \cite{PM12}, the above networks are sustained when a feedback condition and a marginal stability criterion are simultaneously satisfied. The former depends on the maximum energy of the proton distribution and the latter on the density of relativistic protons inside the source. Regardless of the details of the networks involved, the result is an exponential growth of photons, which are produced either directly or indirectly from protons, leading inevitably to proton cooling through photohadronic interactions. The conditions in the source just before the onset of the photon number exponentiation indicate a {\sl critical point} for the system; if the protons inside the source have reached a certain critical density, they become {\sl supercritical}. Any attempt to further increase their density will make the system undergo a {\sl phase transition} by efficiently removing energy from the protons due to photohadronic interactions and transferring it to photons, pairs, and neutrinos. This physical situation is in many ways analogous to ``sand piles'', where just one more grain of sand can cause the sudden collapse of the pile. 

The onset of supercriticality depends on various parameters, such as the magnetic field inside the source, the energy content and the maximum cutoff energy of the protons, and others. In an astrophysical environment one could envisage various ways that lead to supercriticalities. The simplest scenario is to consider that protons, after being accelerated to relativistic energies, are injected inside a confining volume. In the absence of strong losses and fast escape, the injected protons accumulate and their density could become supercritical. The onset of supercriticality (i.e., linear phase of photon outgrowth) can be studied by analytical means as long as the network of the driving physical processes can be isolated and the rate equations simplified \citep{KM92,PM12}. However, the study of the system in the non-linear phase of the photon outgrowth can be followed only with numerical means. Despite the considerable amount of work that has been done so far on various aspects of hadronic supercriticalities (see the references listed above), a systematic search of the relevant parameter space is still missing. 

Here, we numerically explore the phase space of hadronic supercriticalities for a wide range of parameters relevant to non-thermal emitting astrophysical sources. Our goal is to make a comprehensive list of all possible supercriticalities, classify the temporal behaviour of the system, and gain a better understanding of the driving physical processes.

This paper is structured as follows. In Section \ref{Sec:2}, we outline the physical model and the methodology of our numerical investigation, and continue in Section \ref{Sec:3} with a presentation of the parameter space of hadronic supercriticalities. In Section \ref{Sec:4}, we identify the networks of the physical process that drive the hadronic system to the supercritical regime. In Section \ref{Sec:5}, we study the effects of various secondary model parameters on our results. We briefly discuss the astrophysical implications of this study in Section~\ref{Sec:6}.  We continue in Section \ref{Sec:7} with a summary and discussion of our results, and conclude in Section~\ref{Sec:8}. 

\section{Methods}\label{Sec:2}
In order to investigate numerically the phase space of hadronic supercriticalities, we adopt the standard framework of a one-zone radiation model. We consider a spherical source of radius $R$, containing a tangled magnetic field of strength $B$, where pre-accelerated relativistic protons with a power-law energy distribution are uniformly injected at a rate given by
\begin{equation}
\label{eq:inj}
Q_{\rm p}(\gamma_{\rm p},t)=Q_{\rm p,0}\gamma_{\rm p}^{-s} H(\gamma_{\rm p}-\gamma_{\rm p,min})H(\gamma_{\rm p,max}-\gamma_{\rm p})H(t)\ ,
\end{equation}
where $\gamma_{\rm p,min}$ and $\gamma_{\rm p,max}$ correspond to the minimum and maximum proton Lorentz factors, respectively, $s$ is the power-law index of the proton distribution and $H(x)$ is the Heaviside step function. The proton injection compactness $\ell_{\rm p}$
is defined as 
\begin{equation}\label{eq:qp}
\ell_{\rm p}=\frac{\sth L_{\rm p}}{4\pi R m_{\rm p} c^3} =\frac{1}{3} Q_{\rm p,0}t_{\rm cr} \sigma_{\rm T} R \int_{\gamma_{\rm p,min}}^{\gamma_{\rm p,max}}\gamma_{\rm p} ^{-s+1} \rm{d}\gamma_{\rm p}
\end{equation}
where $L_{\rm p}$ is the proton injection luminosity, $t_{\rm cr}=R/c$ is the source light-crossing timescale, $c$ is the speed of light, and $\sigma_{\rm T}$ is the Thomson cross section. Similarly, we define the magnetic field compactness, which is a dimensionless measure of the magnetic energy density $U_{\rm B}$
\begin{equation} 
\label{eq:lb}
\ell_{\rm B}=\sigma_{\rm T}R \frac{U_{\rm B}} {m_{\rm e}c^2},
\end{equation}
where $m_{\rm e}$ is the electron rest mass. The photon compactness is accordingly defined as
\begin{equation}
\ell_{\rm \gamma}=\frac{L_{\rm \gamma} \sigma_{\rm T}}{4\pi R m_{\rm e} c^3} = \sigma_{\rm T} R \frac{U_{\gamma}}{3 m_{\rm e}c^2}\left(1+\frac{\tau_{\rm T}}{3} f(\varepsilon)\right)^{-1},
\label{eq:lgamma}
\end{equation}
where $L_{\rm \gamma}$ and $U_{\gamma}$ are the bolometric photon luminosity and energy density, respectively, $\tau_{\rm T}$ is the Thomson optical depth, and $f(\varepsilon)$ is a function of the photon's energy $\varepsilon$ in units of $m_{\rm e} c^2$ \citep[for the full expression, see][]{MK95}. 

Upon entering the source, the relativistic protons interact with the magnetic field and any soft photons present,
creating radiation and secondaries. We do not consider the presence of external photon
fields, as this would introduce more free parameters, i.e., we make the implicit assumption that the internal photon energy density is always dominant. For the same reason, we do not consider, until Section \ref{Sec:5}, the injection of an accelerated electron (primary) population in order to study the supercriticalities of a {\sl pure hadronic} model first. 

We assume that pions and muons decay instantaneously into secondary particles, and that neutrons do not interact with soft photons before they escape the source (i.e., the source is optically thin to neutron-photon interactions). Neutrinos escape the source without any interactions at the light-crossing time. Thus, at any given time, there are three stable particle species in the source, namely protons, photons, and electron-positron pairs (henceforth, electrons). The temporal evolution of their distributions can be described by three coupled kinetic equations: 
\begin{equation}\label{eq:ke}
\frac{\partial n_{\rm i}}{\partial t}+\frac{n_{\rm i}}{t_{\rm i,esc}}+\mathcal{L}_{\rm i}=Q_{\rm i},
\end{equation}
where $n_{\rm i}$ is the differential number density of particle species $i$, and $t_{\rm i,esc}$ is the respective escape timescale (for photons, $t_{\rm \gamma,esc}=t_{\rm cr}=R/c$).
The operators $Q_{\rm i}$ and $\mathcal{L}_{\rm i}$ denote the injection (source) and loss (sink) terms, respectively. These terms include the following processes: synchrotron radiation for both electrons and protons, photopair (Bethe-Heitler) production, photopion production, photon-photon ($\gamma \gamma$) pair production, inverse Compton scattering, synchrotron self absorption, pair annihilation, and photon down-scattering on cold electrons \citep[for details, see][]{MK05,DMPM12}.

The free parameters of the model are: the radius of the source $R$, the magnetic field $B$, the proton injection compactness $\ell_{\rm p}$ (or, equivalently $Q_{\rm p,0}$), the minimum and maximum proton Lorentz factors ($\gamma_{\rm {p,min}}$ and $\gamma_{\rm{p,max}}$, respectively), the proton and electron escape timescales ($t_{\rm p,esc}$
and $t_{\rm e,esc}$, respectively), and the power-law index $s$.
In order to reduce the number of free parameters we set  $\gamma_{\rm p,\min}=1$, 
and $t_{\rm e,esc}=t_{\rm cr}$ and we keep them constant for all runs. Also, for the next sections we set $t_{\rm p,esc}=1000\, t_{\rm cr}$ and 
$s=2$, and discuss the implications of this choice in Section~\ref{Sec:5}.

For each parameter set (i.e., $R$, $B$, $\gamma_{\rm p, \max}$, and $\ell_{\rm p}$) we solve numerically equation~(\ref{eq:ke}) by utilizing the code of \cite{MK95}, with improved production rates and emissivities introduced by \cite{MK97,MK05}. We use as our initial conditions $n_{\rm i}(\gamma_{\rm i},t=0)=0$ for all species and let the system evolve until some time $t_{\rm end}$, which is taken to be much larger than all other relevant timescales of the system, except when noted.

\section{The phase space of hadronic supercriticalities}\label{Sec:3}
In this section, we use the numerical code described above to perform a systematic search of the parameter space of  hadronic supercriticalities and study the temporal behaviour of a hadronic system in this regime.
In Section~\ref{Subsec:3.1}, we show a typical example of the way a hadronic system moves from the subcritical to the supercritical regime by performing consecutive runs of increasing injection proton luminosity. In Section~\ref{Subsec:3.2}, we perform a search for supercriticalities for various combinations of the source radius $R$ and its magnetic field strength $B$, which can be of relevance to astrophysical sources. Finally, in Section~\ref{Subsec:3.3}, we investigate the role of the maximum proton energy in the development of the supercriticality. 

\subsection{Transition to supercriticality: the role of the proton injection compactness}\label{Subsec:3.1}
\begin{figure}
 \centering
\includegraphics[width=\linewidth]{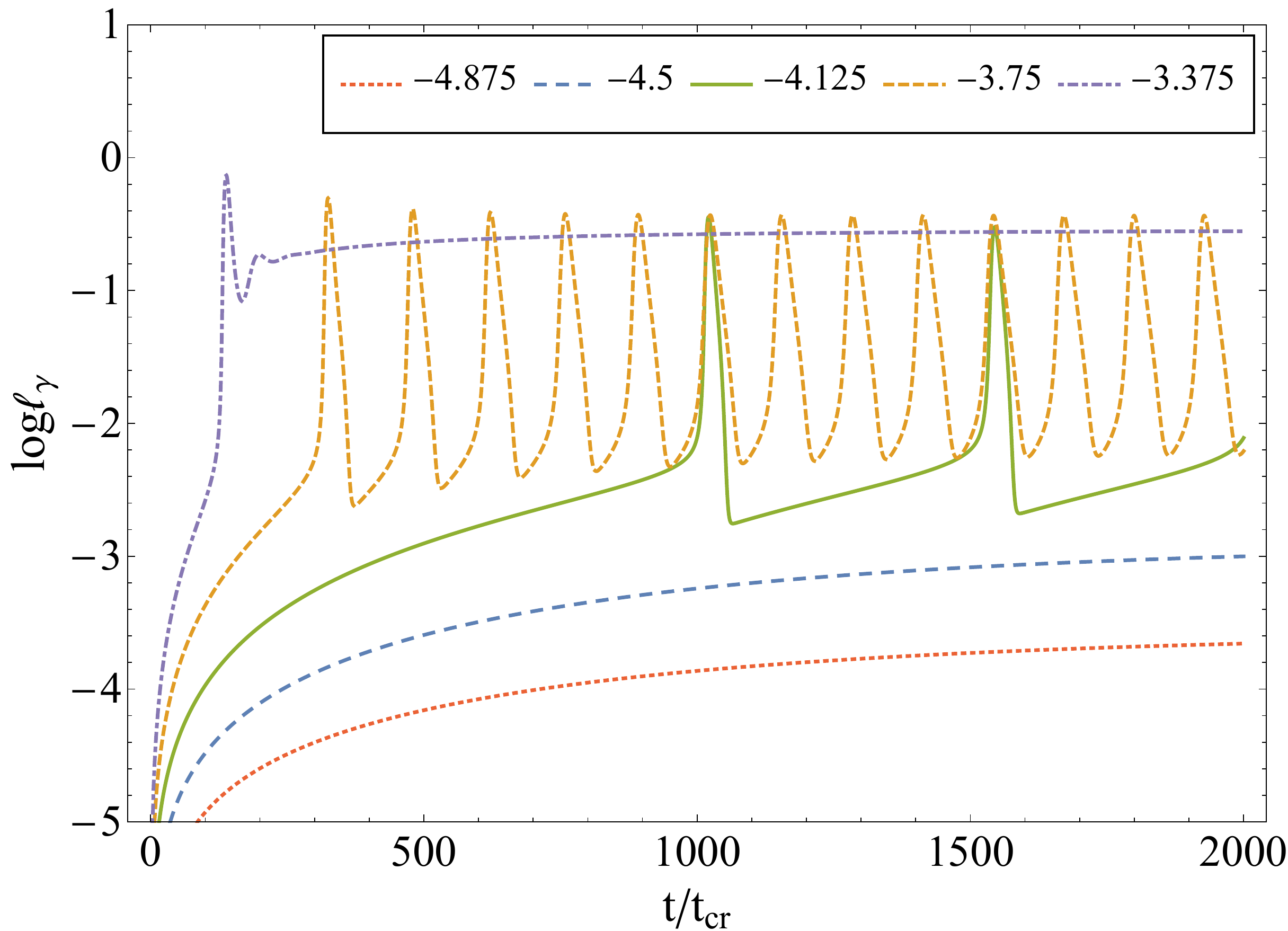}
\caption{The photon compactness (in logarithm) versus time (in units of the light-crossing time $R/c$) for a sequence of $\log\ell_{\rm p}$ values that differ by a factor of $0.375$ (see inset legend). 
Other parameters used here are: $R=10^{15}$~cm, $B=10^{1.5}$~G, $\gamma_{\rm p,max}=10^{6.5}$ and $t_{\rm p,esc}=10^3 t_{\rm cr}$.}
\label{Fig:1}
\end{figure}

We begin our numerical analysis by showing a  representative case of the transition to supercriticality. Here, we perform consecutive runs for increasing values of $\ell_{\rm p}$, as illustrated in Fig.~\ref{Fig:1}, for $R=10^{15}$~cm, $B=10^{1.5}$~G, $\gpmax=10^{6.5}$, and $t_{\rm p,esc}=10^3 \ t_{\rm cr}$.
A similar example was also presented in \cite{PM18}, but for the sake of completeness, we outline the procedure here, as it is central for the understanding of this paper.

\begin{figure}
 \centering
\includegraphics[width=\linewidth]{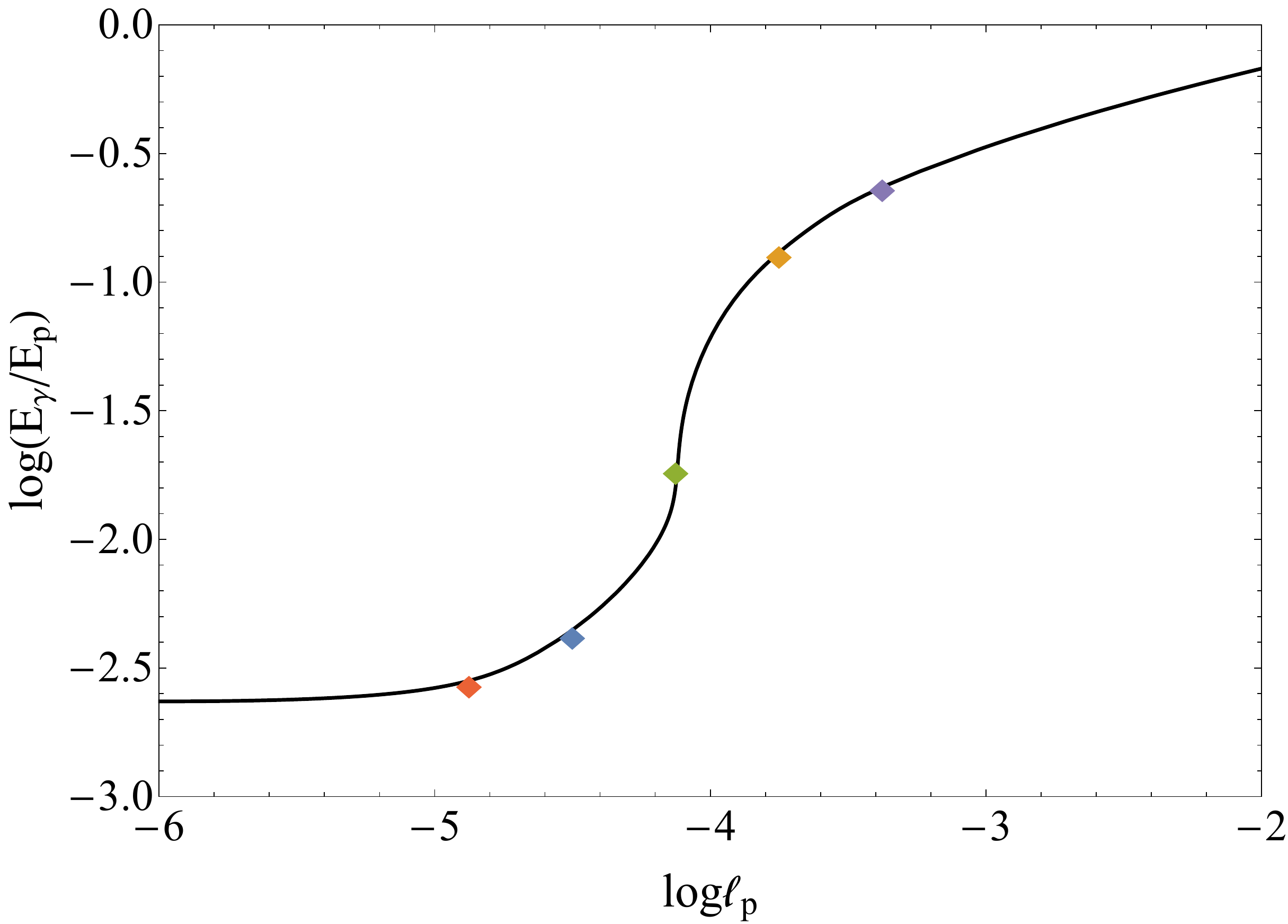}
\caption{Log-log plot of the radiative efficiency, namely the ratio of the total radiated energy in photons, $E_\gamma$, to the total injected energy in relativistic protons, $E_{\rm p}$, as a function of the proton injection compactness $\ell_{\rm p}$ for  $t_{\rm end}=1200 \, t_{\rm cr}$.
Filled diamonds indicate the values of $\log\lp$ used to compute the light curves in Fig.~\ref{Fig:1} (same colour coding used).}  
\label{Fig:2a}
\end{figure} 

\begin{figure*}
 \centering
\includegraphics[width=\linewidth]{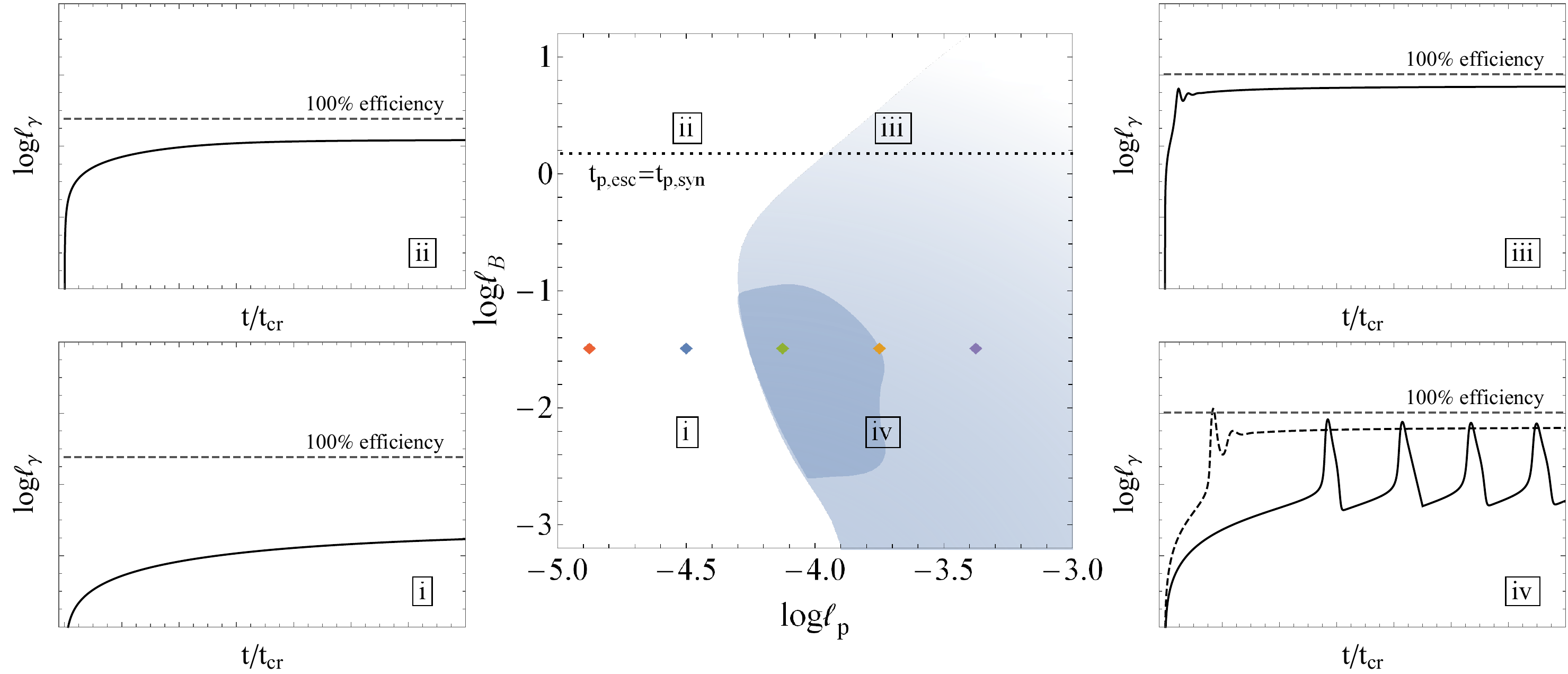}
\caption{The $\ell_{\rm B}-\ell_{\rm p}$ parameter space of a pure hadronic system (central plot) with $R=10^{15}$~cm, $\gamma_{\rm p,max}=10^{6.5}$, and 
$t_{\rm p,esc}=10^3 \, t_{\rm cr}$.
The temporal evolution of the system varies across the $\ell_{\rm B}-\ell_{\rm p}$ parameter space, as illustrated by the four light curves in the side panels. The dark coloured area corresponds to values where the system exhibits limit cycles (see solid line in panel iv), while the light coloured area corresponds to violent relaxations (see dashed line in panel iv).
The gradient indicates the progressive transition from limit-cycle behaviour to a high photon compactness steady-state one (see panel iii). The dotted line indicates the locus of points where $t_{\rm p,esc}=t_{\rm p,syn}(\gamma_{\rm p,max})$, while filled diamonds mark the cases shown in Fig.~\ref{Fig:1} (same colour coding used).  
}
\label{Fig:2}
\end{figure*}  

For low values of $\lp$ (dotted red and dashed blue lines) the system is still in the subcritical regime and it reaches a steady state, typically after a few hundred crossing times. The photon emission, whose light curve is shown in the figure, is mostly the result of proton synchrotron radiation and it increases linearly with $\lp$.
As $\lp$ increases, the proton number density exceeds some critical value, which will be defined in Section~\ref{Subsec:3.2},
and the system is driven to supercriticality. This transition is manifested with a sudden increase in the radiative efficiency (see Fig.~\ref{Fig:2a} and relevant discussion in next paragraph) and bursty light curves. The bursts may occur (quasi)periodically (limit-cycle behaviour; see solid green and dashed orange lines) with a period that depends on $\lp$. More specifically, higher values of $\lp$ lead to more frequent outbursts, while the first burst progressively appears at earlier times.
For even higher values of $\lp$, the number of bursts in the light curve reduces to one before the photon compactness reaches a constant value
(dash-dotted purple line); henceforth, we refer to such cases as violent relaxations.
This is essentially a high-efficiency steady-state regime, where the photon emission is produced via photohadronic interactions on the proton-synchrotron radiation. Therefore, we can summarize the general behaviour of the system with increasing $\ell_{\rm p}$ as: subcritical (low steady state) $\rightarrow$ supercritical (limit cycles) $\rightarrow$ supercritical (violent relaxation).

The characteristic supercritical behaviour of the photon light curve \citep[see also][]{Stern91, MK05} can be explained from the fact that once the protons enter the supercritical regime, photons grow exponentially with $n_\gamma\propto e^{\alpha t}$ ($\alpha > 0$) until they abruptly cool the protons. Therefore, photons reach eventually a maximum density before they escape the source, thus producing a characteristic outburst. The index $\alpha$ is a function of the system's parameters (e.g., magnetic field strength, source radius, and others). The higher its value, the stronger the supercriticality. 

\begin{figure*}
 \centering
 \includegraphics[width=0.92\linewidth]{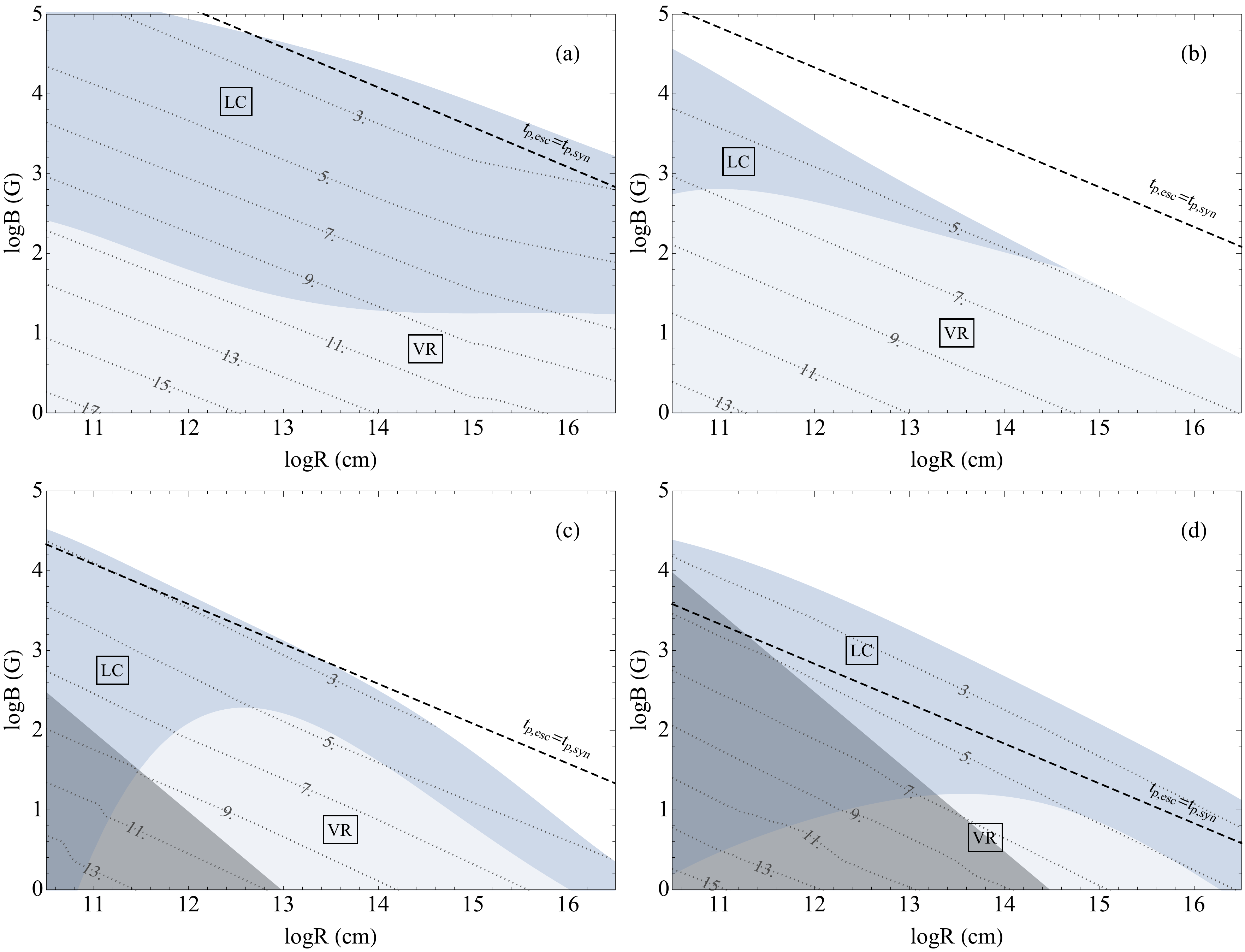} \caption{The $B-R$ parameter space of hadronic supercriticalities for fixed maximum proton Lorentz factors: (a) $\gamma_{\rm p,max}=10^{3.5}$, (b) $\gamma_{\rm p,max}=10^{5}$, (c) $\gamma_{\rm p,max}=10^{6.5}$, (d) $\gamma_{\rm p,max}=10^{8}$. The dark coloured region corresponds to limit cycles (LC) and the light coloured region to violent relaxations  (VR). The critical proton compactness cannot be defined for parameters drawn from the white region, as no outburst appears at $t < t_{\rm p,esc}$. 
  Contours of the ratio $U_{\rm p,crit} /U_{\rm B}$ (in logarithmic scale) are overplotted (dotted lines). The darkest coloured areas (bottom left corner in panels c and d) correspond to proton gyroradii $r_{\rm g}>R$ and are thereby forbidden. The dashed line corresponds to $t_{\rm p,syn}(\gamma_{\rm p,max})=t_{\rm p,esc}$ where $t_{\rm p,syn}(\gamma_{\rm p,max})$ is the synchrotron cooling timescale of the maximum energy protons.}
 \label{Fig:3}
\end{figure*}

A result of the transition to the supercriticality is the increase of the radiative efficiency of the system. This is illustrated in Fig.~\ref{Fig:2a}, which shows the ratio of the total radiated energy in photons, $E_\gamma$, to the total injected energy in relativistic protons, $E_{\rm p}$, as a function of the proton injection compactness $\ell_{\rm p}$ for the parameters used in Fig.~\ref{Fig:1} (same colour coding used) and  $t_{\rm end}=1200 \, t_{\rm cr}$. 
All runs for low values of $\ell_{\rm p}$ (e.g., $\ell_{\rm p}\sim 10^{-6}-10^{-5}$) correspond to low-efficiency steady states dominated by proton synchrotron radiation (subcritical regime). As $\ell_{\rm p}$ increases  ($\ell_{\rm p}\gtrsim10^{-5}$), photohadronic interactions become progressively more important in both photon production and proton cooling,  
since these processes depend on the product of the number density of protons and photons\footnote{There is an analogy here with the synchrotron self-Compton process of leptonic plasmas.}.
This results to an increase of the system's radiative efficiency, however the behaviour is still subcritical. 
The phase transition to supercriticality occurs with
the appearance of even one outburst
($ \ell_{\rm p}$ values around green point).
This results in a dramatic increase of the efficiency  as a large fraction of the energy stored in protons is burned explosively. As a result, the efficiency curve turns abruptly upwards, reaching values close to 0.1 (entrance to supercriticality). The increase in the efficiency continues during the whole limit cycle phase 
and only when the system reaches the violent relaxation phase 
it starts moving asymptotically to efficiencies close to unity. Because of the neutrino emission, the photon efficiency will never become exactly one, even if the protons radiate all their energy away. In short, Fig.~\ref{Fig:2a} shows a representative curve for a hadronic system's radiative efficiency as this transits from the subcritical regime to supercriticality. Although the actual values for the efficiency and critical proton compactness depend on the chosen parameters, 
the overall shape is retained for a wide range of initial parameters.

While the previous example shows a rather typical way to enter the supercritical regime, this is by no means the only one. This is illustrated in Fig.~\ref{Fig:2}, where we show the $\ell_{\rm B}-\lp$ parameter space of a hadronic system for the same parameters as in Fig.~\ref{Fig:1}, i.e., $R=10^{15}$~cm,  $\gamma_{\rm p,max}=10^{6.5}$ and $t_{\rm p,esc}=10^3 t_{\rm cr}$. 
For comparison purposes, we also indicate with filled diamonds the parameters used for the runs in Fig.~\ref{Fig:1}.
 One can identify four regions of interest in the $\ell_{\rm B}-\lp$ plane. For both low $\lp$ and $\lB$ values (region i) the system is subcritical and protons are inefficient in radiating away their energy. For higher $\lB$ values (and low values of $\lp$) the system is still subcritical, but proton synchrotron cooling and synchrotron radiation become efficient (region ii). The supercritical regime (depicted as grey area) occupies the right-hand side of the $\ell_{\rm B}-\lp$ plane (regions iii and iv), i.e., it requires high values of $\lp$ to allow the proton number density to build up to the required critical values. The dark grey region is where limit cycles appear
 while lighter grey areas correspond to supercritical solutions that end up eventually in a steady state through violent relaxations  (see top and bottom right panels)\footnote{The system's temporal behaviour can be understood qualitatively with analytical means  
-- see Appendix \ref{App:B}.}. Finally,
region iii of the parameter space is characterized by both fast proton  cooling (synchrotron and photohadronic), and by an abrupt transition to a high-efficiency steady state. Typically, this transition occurs very quickly after the onset of proton injection and its exact time behaviour depends on the competition between the two cooling processes.

\subsection[The B-R phase space]{The $B-R$ phase space}\label{Subsec:3.2}
So far, we have discussed the system's transition to supercriticality for fixed $R$ and $\gamma_{\rm p,max}$. We can extend the parameter space search even further by looking for supercriticalities in the  $B-R-\gamma_{\rm p,max}$ phase space. To do so, we have to define the minimum proton injection compactness required for the transition to supercriticality.

We therefore define as \textit{critical proton compactness} ($\ell_{\rm p,crit}$) the value of $\ell_{\rm p}$ that produces a photon flare at $t_{\rm app}=t_{\rm p,esc}$, where $t_{\rm app}$ is the appearance time of the first peak in the light curve. According to this definition, the green symbol in Figs.~\ref{Fig:1}-\ref{Fig:2} indicates $\ell_{\rm p,crit}$. However, a transition to supercriticality (grey-coloured region) can also occur for slightly lower values of $\ell_{\rm p}$. These values can still lead to the production of bursts of radiation but at longer $t_{\rm app}$. If no outburst is produced, which is the case for region ii (and sometimes region iii) in Fig.~\ref{Fig:2}, $\ell_{\rm p, crit}$ cannot be defined. Regardless, our definition of $\ell_{\rm p,crit}$ is sufficient for the parameter space search that follows.

For each $\ell_{\rm p,crit}$ value, we also compute the corresponding proton energy density $U_{\rm p,crit}$, which is
the maximum density that the protons acquire before the first photon outburst occurs, and the ratio $U_{\rm p,crit}/U_{\rm B}$, which is a measure of the deviation from energy equipartition at the onset of supercriticality.
Our results are summarised in Fig.~\ref{Fig:3}, which shows the logarithmic $B-R$ phase space of hadronic supercriticalities for four values of $\gamma_{\rm p,max}$ (for the parameter ranges employed, see Table~\ref{Table:1}).

\begin{table}
\centering
\begin{tabular}{ccc}
\hline
Physical quantity & Minimum value & Maximum value \\ \hline
R (cm) & $10^{11}$ & $10^{16}$ \\
B (G) & $10^{0}$ & $10^{5}$ \\
$\gamma_{\rm p, max}$ & $10^{3}$ & $10^{9}$ \\ \hline
\end{tabular}%
\caption{Range of values of the radius $R$, magnetic field strength $B$, and Lorentz factor $\gamma_{\rm p, max}$ used throughout the paper.}
\label{Table:1}
\end{table}

In all panels, one can identify two regions of interest:
\begin{itemize}
 \item The upper right corner of the phase space (white region) corresponds to the fast proton synchrotron cooling regime, where protons radiate efficiently their energy via synchrotron. The system always reaches a steady state irrespective of the value of $\ell_{\rm p}$ (see also regimes ii and iii in Fig.~\ref{Fig:2}). In this region, the critical proton compactness cannot be defined, as no outburst occurs at $t<t_{\rm p, esc}$.
 \item The lower left region (various shades of grey) corresponds to the slow synchrotron cooling regime, where a transition to supercriticality is always possible. Different shades are used to denote different manifestations of  supercriticality: the (quasi-)periodic outbursts or limit cycles (denoted as (LC) in the plots) are depicted in dark grey, while the violent relaxations (VR) to a steady state are shown with light grey (see also regime iv in Fig.~\ref{Fig:2}). For $B$, $R$ values drawn from the lower left dark-coloured region, the gyroradius of the highest energy protons, i.e., $r_{\rm g}=\gamma_{\rm p,max} m_{\rm p}c / e B$, is larger than the assumed source radius \citep{Hillas}. Thus, this part of the parameter space is physically forbidden.
\end{itemize}

We generally find that for an increasing $R$ the manifestation of supercriticality changes, as limit cycles give their place to violent relaxations, especially for large values of $\gamma_{\rm p,max}$. This is due to the fact that the conditions for supercriticality depend on the photon  compactness which is related to the size of the source (see equation \ref{eq:lgamma}). 

Frequent limit cycles with large-amplitude photon outbursts appear both for small and large values of $\gamma_{\rm p,max}$, while this behaviour becomes weaker, by showing violent relaxations, for intermediate values of this parameter.
This rather complicated dependence of  supercriticality's manifestations on the maximum proton energy is explained in Section~\ref{Sec:4}.

The dependence of the supercriticality on $B$ is also complicated. However, there is an overall tendency, as $B$ increases and $R$ is kept fixed, for the system to successively transit from a violent relaxation to a limit-cycle behaviour and, finally, to steady state (i.e., from region iv to region iii in Fig.~\ref{Fig:2}). Note also that in all panels we have drawn the line $t_{\rm p,syn}(\gamma_{\rm p,max})=t_{\rm p,esc}$, where $t_{\rm p,syn}(\gamma_{\rm p,max})$ is the synchrotron cooling timescale of the maximum energy protons. This line, however, does not describe very accurately the transition between the sub- and supercritical regimes and it has to be taken more as an indication.

The transition to supercriticality requires generally higher energy densities in protons than in magnetic fields. The ratio $U_{\rm p,crit}/U_{\rm B}$ decreases for increasing magnetic field strengths (see contours in all four panels). Although this trend is generic, the exact values of the ratio $U_{\rm p,crit}/U_{\rm B}$ depend on the maximum proton energy $\gamma_{\rm p,max}$  as well as on the power-law index $s$ (see Section \ref{Sec:5:2}). 

Summarizing, our parameter exploration showed that supercriticalities may appear for  a wide range of $B, R$ and $\gamma_{\rm p,max}$ values, thus making them potentially relevant for all types of compact astrophysical objects (for more details, see Section~\ref{Sec:6}).

\begin{figure}
 \centering
 \includegraphics[width=\linewidth]{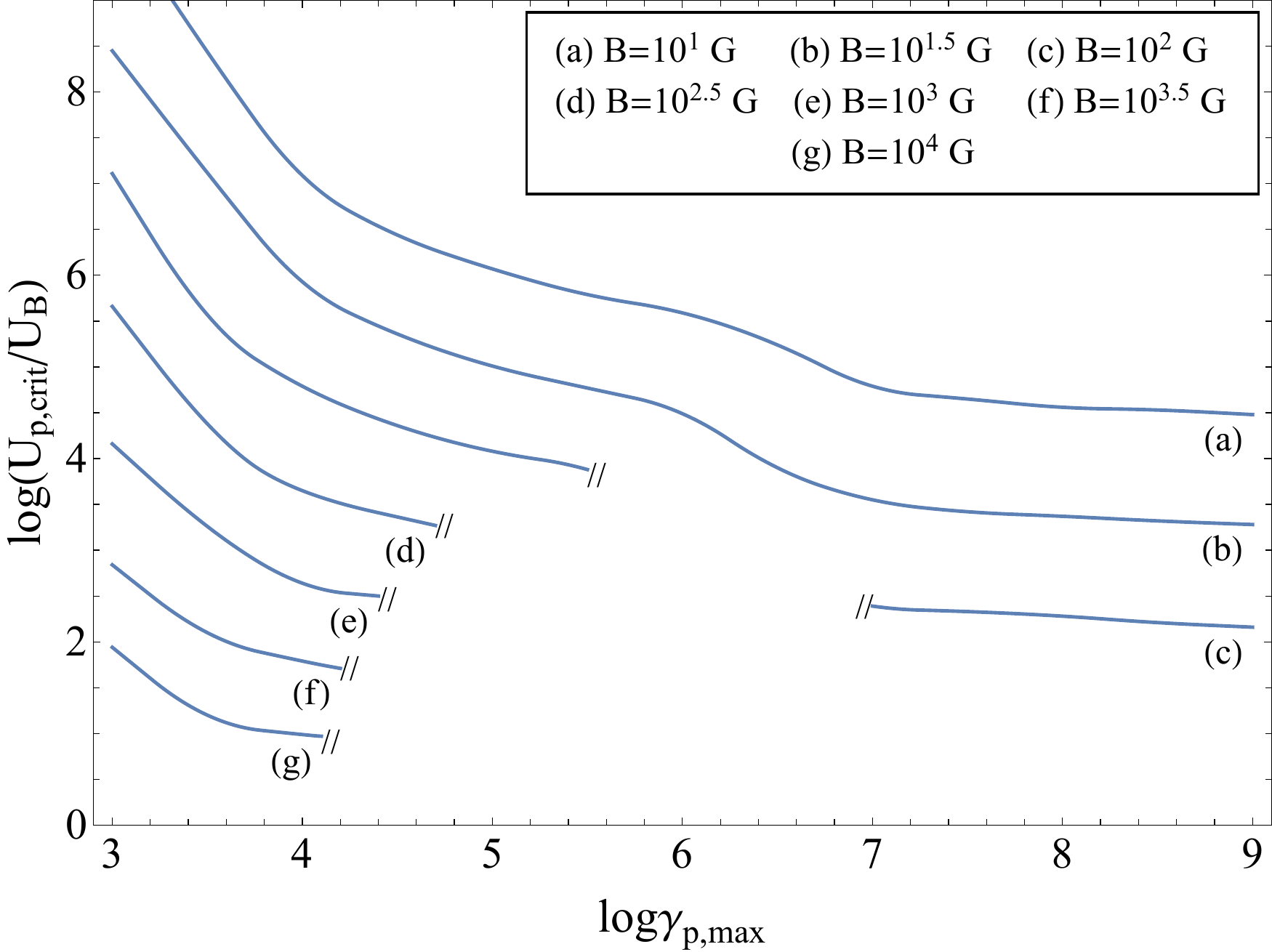}
 \caption{The ratio of the critical proton energy density $U_{\rm p,crit} $ to the magnetic energy density $U_{\rm B}$ plotted as a function of $\gamma_{\rm p, \max}$ for a source with radius $R=10^{15}$~cm and different values of the magnetic field (see inset legend). The curves are truncated whenever $U_{\rm p, crit}$ cannot be defined (i.e., the system reaches a steady state without showing any photon outburst). }
 \label{Fig:4}
\end{figure}

\subsection{The role of the maximum proton energy} \label{Subsec:3.3}

We next turn on investigating in more detail the effects of the maximum proton energy $\gpmax$ on the critical proton energy density $U_{\rm p,crit}$. For this purpose, we search for the onset of supercriticality as a function of $\gamma_{\rm p,max}$.
Our results are presented in Fig.~\ref{Fig:4} where the ratio $U_{\rm p,crit}/U_{\rm B}$ is plotted as a function of $\gamma_{\rm p,max}$ for a source of radius $R=10^{15}$~cm and different values of the magnetic field $B$ ranging from $B=10$~G up to $10^{4}$~G with an increment of 0.5 in logarithm. 
 Overall, we find that the critical proton energy density $U_{\rm p,crit}$ decreases for higher values of $\gamma_{\rm p,max}$. There is a strong dependence of $U_{\rm p, crit}$ on $\gamma_{\rm p,max} \lesssim 10^4$, but the dependency weakens for high values (i.e., $\gamma_{\rm p,max} \gtrsim 10^7$). For $\gamma_{\rm p,max} < 10^3$, it is becoming increasingly difficult to transit to the supercritical regime,
and for all practical purposes one can safely assume that no supercriticality occurs below this value. Generally, as $B$ increases (i.e., moving from curves a to g),
we find that the system reaches a steady state without showing any outbursts for a wide range of $\gamma_{\rm p, max}$ values. The concept of a critical proton energy density becomes irrelevant for the entire lower part of the diagram (i.e., below curve c), which is intentionally left blank.

\section{The physical mechanisms of supercriticality}\label{Sec:4}
In the previous sections, we showed that, for a wide range of values for the main physical parameters ($B$, $R$, and $\gpmax$), there exists a critical value of the proton compactness which drives the system into supercriticality. In this section, we identify the physical processes that play a dominant role in this transition.

As a guide, we will use the analytical results of \cite{KM92} and \cite{PM12} -- henceforth KM92 and PM12, respectively. As it was demonstrated in these studies, hadronic supercriticalities arise as a result of specific networks of physical processes (or feedback loops). In what follows, we briefly present the basic ideas behind these papers before applying them to this study's findings.

\subsection{Photopair or Photopion - Electron Synchrotron Loop}\label{Subsec:4.1}
The premise behind the photopair-electron synchrotron ($PeS$) and the photopion-electron synchrotron ($P \pi S$) loops is fairly simple:
electron-positron pairs that are created directly in photopair or indirectly (through the decay of charged mesons) in photopion production processes radiate synchrotron photons. These, in turn, can serve as targets for additional photopair or photopion interactions of protons, thus leading to the production of more pairs, which will radiate via synchrotron and so on. Thus, a loop of physical processes describing the exponentiation of pairs and photons is formed.

KM92 demonstrated that both the $PeS$ and $P \pi S$ loops
can lead the system to supercriticality and derived, using simple analytical formulas, the threshold and the marginal stability criteria that are required to make the above feedback loops operate. In the case of the $PeS$ loop, the feedback criterion is that the proton Lorentz factor should exceed a critical value  given by:
\begin{equation}
\gamma_{\rm p,crit}^{\rm PeS}=\left (\frac{2}{b} \frac{m_{\rm p}}{m_{\rm e}}\right)^{1/3}
\label{gBHcr}
\end{equation}
while the corresponding value in the case of the $P \pi S$ loop is{\footnote {This relation has not been derived explicitly in KM92.}}
\begin{equation}
\gamma_{\rm p,crit}^{\rm P\pi S}=\left(\frac{1}{b\eta_{\pi^\pm}^2}\frac{m_{\pi}}{m_{\rm e} } \left(1+\frac{m_{\pi}}{2m_{\rm p}}\right)  \right)^{1/3}    
\label{gpicr}
\end{equation}
where $m_{\rm p}$, $m_{\rm e}$, and $m_{\rm \pi}$ are the proton, electron and pion masses, respectively, $b=B/B_{\rm crit}$, $B_{\rm crit}=4.4\times10^{13}$~G is the Schwinger magnetic field, and $\eta_{\rm \pi^\pm} \simeq 150$ relates, on average, the Lorentz factor of the secondary pairs produced from charged pion decay to the Lorentz factor of the parent proton \citep{DMPM12}. By comparing the above relations, we find that $\gamma_{\rm p,crit}^{\rm P\pi S} < \gamma_{\rm p,crit}^{\rm PeS}$ is always satisfied.

These critical values of the proton Lorentz factor are derived by requiring that the protons are energetic enough to pair-produce or pion-produce on the synchrotron photons emitted by their secondaries.  Both expressions are approximate as  they were derived using $\delta-$functions for the electron synchrotron emissivity (i.e.,  the relation $x_{\rm syn}=b\gamma^2$ was used) and for the production spectra of the secondaries. However, as we will show later, even under these rough approximations, these relations can qualitatively describe the behaviour of the system for low $\gamma_{\rm p,max}$. 

 The marginal stability criterion
basically requires that {the proton density} is high enough so that the synchrotron photons radiated from the secondaries can produce either more pairs ({\sl PeS}) or more pions ({\sl P$\pi$S}) before they escape from the source.  Because the derivation of the marginal proton number density $n_{\rm p,marg}$ is less straightforward than the one for the critical proton Lorentz factor, we simply repeat the finding of KM92, namely $n_{\rm p,marg}^{\rm P\pi S} > n_{\rm p,marg}^{\rm PeS}$. This means that if the feedback criterion is satisfied for both loops, then the {\sl PeS} loop occurs more easily as it requires lower proton densities.

\subsection{Photo-Quenching Loop}\label{Subsec:4.2}

PM12 examined another feedback loop relevant in hadronic systems
--  we will use the abbreviation $\gamma Q$ -- that was based on the instability of automatic photon quenching \citep{Stawarz07,PM11}.  This describes essentially the spontaneous outgrowth of electron-positron pairs and their synchrotron radiation in a source when its $\gamma-$ray compactness exceeds some critical value and, as such, it does not necessarily involve protons. 
\citet{PM11} showed that there is a regime in the diagram of the differential radiated luminosity versus photon energy that is ``forbidden'' in the sense that $\gamma-$rays cannot exist in steady state there. If $\gamma-$rays enter this regime they become self-quenched as they spontaneously produce soft photons which, in turn, absorb them. This process is, in many ways, analogous to the loops discussed in the previous section with $\gamma-$rays replacing protons and $\gamma\gamma$ absorption replacing photopair or photopion production. The feedback criterion for automatic photon quenching translates into a critical value for the photon energy
\begin{equation}
 \epsilon_{\gamma,{\rm crit}}=2 b^{-1/3}
\end{equation}
while the marginal stability  criterion sets a critical value for the $\gamma$-ray compactness, which can be written approximately as (see equation~34 of \citealt{PM11})
\begin{equation}
\ell_{\rm \gamma ,marg}^{\rm inj}=\frac{ 3b^{2}\epsilon_{\rm \gamma}^{2}}{8}\left [ \left ( \frac{b \epsilon_{\rm \gamma}}{2} \right )^{3/2} - \frac{8}{\epsilon_{\rm \gamma}^{3}} \right ]^{-1}.
\label{eq9}
\end{equation}

In this study, we are seeking a proton-related mechanism that is, directly or indirectly, producing $\gamma-$rays that simultaneously satisfy the above criteria. These mechanisms are summarised below.
\begin{itemize}
 \item {Proton synchrotron radiation. It can produce directly the required $\gamma-$rays. }
 \item{Photopair production process. It can produce $\gamma-$rays indirectly through the synchrotron/ICS radiation of its secondary electron-positron pairs.} 
 \item {Photopion production process. It can produce $\gamma-$rays either directly through the channel $\pi^0\rightarrow 2\gamma$ or indirectly through the synchrotron/ICS radiation of the secondary electron-positron pairs produced in $\pi^\pm\rightarrow\mu^\pm\rightarrow e^\pm$.}
\end{itemize}
Out of the aforementioned processes, only
proton synchrotron requires rather extreme conditions to meet the threshold condition \citep[i.e., very large values of $B$ and $\gamma_{\rm p,max}$,][]{14MPGM} --
see Appendix  \ref{App:A} for  estimates on the required thresholds of all processes. In the next subsection, we investigate all the aforementioned loops with the help of the numerical code described in Section~\ref{Sec:2}.

\begin{figure*}
 \centering
 \includegraphics[width=\linewidth]{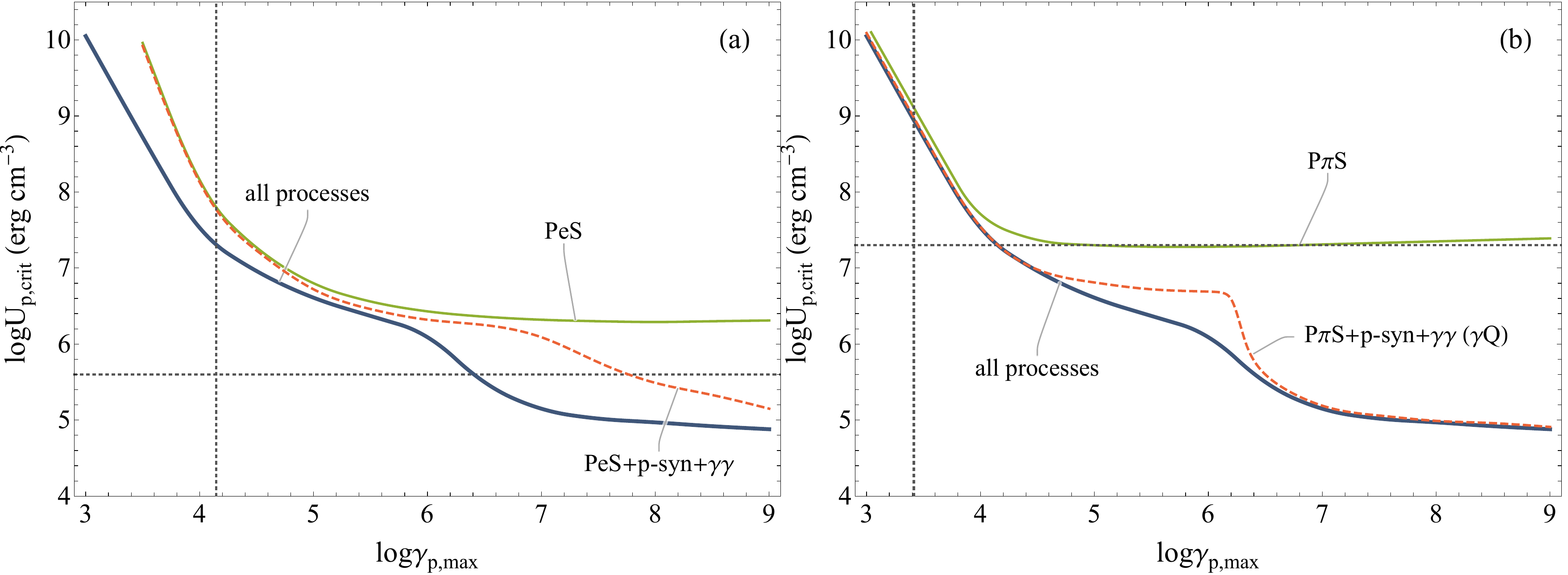}
 \caption{ Plot of the critical energy density $U_{\rm p,crit}$ versus $\gamma_{\rm p,max}$ for $R=10^{15}$~cm and $B=10^{1.5}$~G. The thick solid blue line is computed when all processes are used in the numerical code. The coloured thin lines correspond to combinations of processes that contribute to supercriticality including photopair (Bethe-Heitler) production (panel a) and photopion production (panel b). Dotted lines plot the analytical estimations of KM92.
 }
 \label{Fig:5}
\end{figure*}
\subsection[The Up,crit-γp,max phase space]{The $U_{\rm p,crit}-\gamma_{\rm p,max}$ phase space}

Using the code and switching off all processes except those processes that are key for the {\sl P$\pi$S} and {\sl PeS} loops, we compute  $U_{\rm p,crit}$\footnote{This is computed in exactly the same way as when all processes are included in the code.} and plot it against $\gamma_{\rm p,max}$ in Fig.~\ref{Fig:5}.
Panel (a) is using photopair production as its main proton loss mechanism, panel (b) is using photopion production, while both have only electron synchrotron switched on. Here, we have adopted $R=10^{15}$~cm and $B=10^{1.5}$~G, but our results are largely independent of this choice. We find that the combination of only these processes is adequate to produce the phenomenology addressed in the previous sections where the full code was used (e.g., flaring, limit-cycle behaviour, and others). 

The solid green line in panel (a) shows the critical energy density of the {\sl PeS} loop, while the solid green line in  panel (b)  shows the corresponding curve for the {\sl P$\pi$S} loop. 
On the same plot we also show the analytical results of KM92. The vertical dotted lines in panels (a) and (b) indicate $\gamma_{\rm p,crit}^{\rm PeS}$ and $\gamma_{\rm p,crit}^{\rm P \pi S}$ derived from equations ~\ref{gBHcr} and \ref{gpicr}, respectively. 
These have to be compared with the numerically obtained values of $\simeq 3\times10^3$ and $\simeq 10^3$, respectively -- see the ascending green lines in Fig. \ref{Fig:5}.
The use of the full synchrotron emissivity causes the  numerically derived values of $\gamma_{\rm p,crit}$ to fall below
the analytical ones that were derived using $\delta-$functions instead. The horizontal dotted lines (in both panels) show the marginal proton energy density (for either feedback loop) derived by KM92 using the marginal stability condition. 
For the  adopted values of $B$ and $R$, these read $U_{\rm p,marg}^{\rm PeS} \simeq 4\times10^5~ \rm erg ~cm^{-3}$ and $U_{\rm p,marg}^{\rm P\pi S} \simeq 2\times10^7 ~\rm erg ~cm^{-3}$ (see equations ~6 and 8 of KM92, respectively).
It is worth mentioning that the numerical results are very close to the analytical ones bearing in mind that $U_{\rm p,crit}$ and $U_{\rm p, marg}$ are not defining the same quantity; $U_{\rm p,crit}$ is the maximum energy density that the protons acquire before a photon outburst occurs, while $U_{\rm p,marg}$ is the minimum energy density required to bring the system to supercriticality. In all cases $U_{\rm p,crit}>U_{\rm p,marg}$.  

The fact that the numerical values of $U_{\rm p,crit}^{\rm PeS}$ and $U_{\rm p,crit}^{\rm P\pi S}$ are very close to the blue line in Fig. \ref{Fig:5},  which is the critical density when all processes are taken into account, suggests that the {\sl PeS} and {\sl P$\pi$S} loops  are the main networks of physical processes behind the supercritical behaviour of the system for $\gamma_{\rm{p,max}}\lesssim 10^6$. Indeed, we find numerically that the addition of proton synchrotron emission or $\gamma\gamma$ pair production to the above loops can only slightly decrease the critical proton energy densities for $\gamma_{\rm{p,max}}\lesssim 10^6$ (compare e.g., dashed orange and solid green lines). However, for high enough values of $\gamma_{\rm p,max}$ (here $\gtrsim 10^6$), we find that $U_{\rm p,crit}$ decreases significantly when both proton synchrotron and $\gamma\gamma$ absorption are taken into account (compare dashed orange and solid green lines in both panels). These results imply that  there is another network of processes driving the system's supercritical behavior at high $\gpmax$ values. This network ($\gamma Q$) relies mainly on $\gamma \gamma$ absorption and synchrotron radiation. 

Inspecting both panels of Fig.~\ref{Fig:5}, one can see the contribution of each of the three basic networks discussed above in the case where all processes are used (blue line). The supercriticality starts operating for values of $\gamma_{\rm p,max}>\gamma_{\rm p,crit}^{\rm P\pi S}$, albeit at high values of $U_{\rm p,crit}$. As $\gamma_{\rm p,max}$ increases, there is a sharp drop of the required critical proton density, and at intermediate values of $\gamma_{\rm p,max}$ it is the $PeS$ loop that is mainly responsible for the onset of the non-linearity. Finally, at high values of $\gamma_{\rm p,max}$ it is the $\gamma Q$ network that takes over as it requires the lowest possible critical density -- in the present example the onset of this loop results in a more than one order of magnitude decrease of $U_{\rm p,crit}$. Furthermore, it can be deduced that the aforementioned three loops cover all the possible ways of producing supercriticalities, i.e., while it might be possible, in principle, for other networks to operate, they can do so at the cost of a higher critical density.

Finally, inspection of Fig.~\ref{Fig:2} together with Fig.~\ref{Fig:5} reveals that the supercriticality is strong (i.e., faster exponential growth of photons) when only one loop operates (e.g., the $\gamma Q$ loop for high enough proton energies). When the parameters are such that two loops operate simultaneously (e.g., the $PeS$ and $\gamma Q$ loops for intermediate values of $\gamma_{\rm p,max}$), their superposition weakens the supercriticality -- compare panels (b) and (d) in Fig.~\ref{Fig:3}.

\subsubsection{Synchrotron cooling}
With the help of the aforementioned results one can find some explanation for the rather perplexing role of proton synchrotron cooling on the manifestation of supercriticality (see Fig.~\ref{Fig:3}). Indeed, if the key physical idea behind hadronic supercriticalities is the build-up of a critical proton density in the source, then panel (d) in Fig.~\ref{Fig:3} cannot be explained, as we find supercritical behaviours even for strong synchrotron cooling (i.e., the grey coloured region extends beyond the dashed line). These results can be understood as follows. Synchrotron cooling should not play a critical role in the manifestation of supercriticality, unless a feedback loop that involves protons directly dominates ($PeS$ or $P \pi S$). Indeed, for low enough $\gpmax$ values (see e.g., panel (a) in Fig.~\ref{Fig:3}), it is expected that either $PeS$ or $P \pi S$ will be the main driver of the supercriticality (see also Fig.~\ref{Fig:5}).  This explains why, in this case, supercriticality is found for a wide range of parameters with $t_{\rm p,syn}<t_{\rm p,esc}$. On the other hand, the $\gamma Q$ loop depends only indirectly on the proton density. Even cooled protons can produce $\gamma-$rays with a compactness greater than the critical value, which is needed for the onset of the $\gamma Q$ loop. Therefore, in parameters regimes where the $\gamma Q$ loop prevails, proton synchrotron cooling does not play such an important role in determining the development of   supercriticalities (e.g., panel (d) in Fig.~\ref{Fig:3}). Even so, when proton synchrotron cooling becomes very fast, then the variability dies out,  and the system always reaches a steady-state (see top right corner in all panels of Fig.~\ref{Fig:3}).

\section{Effects of other parameters} \label{Sec:5}
In this section, we investigate the role of various other parameters and factors that were not included in our calculations so far. 

\subsection{Primary electrons}\label{Sec:5:1}
The calculations presented in the previous sections were based upon the assumption that only relativistic protons were injected in the source. Despite the fact that this assumption helped us define the hadronic supercriticality (Section \ref{Sec:3}) and identify the driving physical networks (Section \ref{Sec:5}), it is not a realistic one because most acceleration mechanisms that could operate in astrophysical sources can, in principle, accelerate both protons and electrons to high energies. Therefore, in this section, we add an injection term in the electron kinetic equation (see equation \ref{eq:ke}), with the objective of studying their effects on hadronic supercriticalities. In order to keep the involved free parameters at a minimum, we assume that the electrons are also accelerated into a power law having the same index as protons. We also assume $\gamma_{\rm e,min}=\gamma_{\rm p,min}=1$ and
$\gamma_{\rm e,max}=\gamma_{\rm p,max}$ -- our results turn out to be rather insensitive to these assumptions. Yet, there is one important parameter, namely the electron injection compactness, $\ell_{\rm e}$, which is defined analogously to the proton compactness with $m_{\rm e}$ replacing $m_{\rm p}$ in equation \ref{eq:qp}, and is a dimensionless measure of the injected electron luminosity.

\begin{figure}
 \centering
 \includegraphics[width=\linewidth]{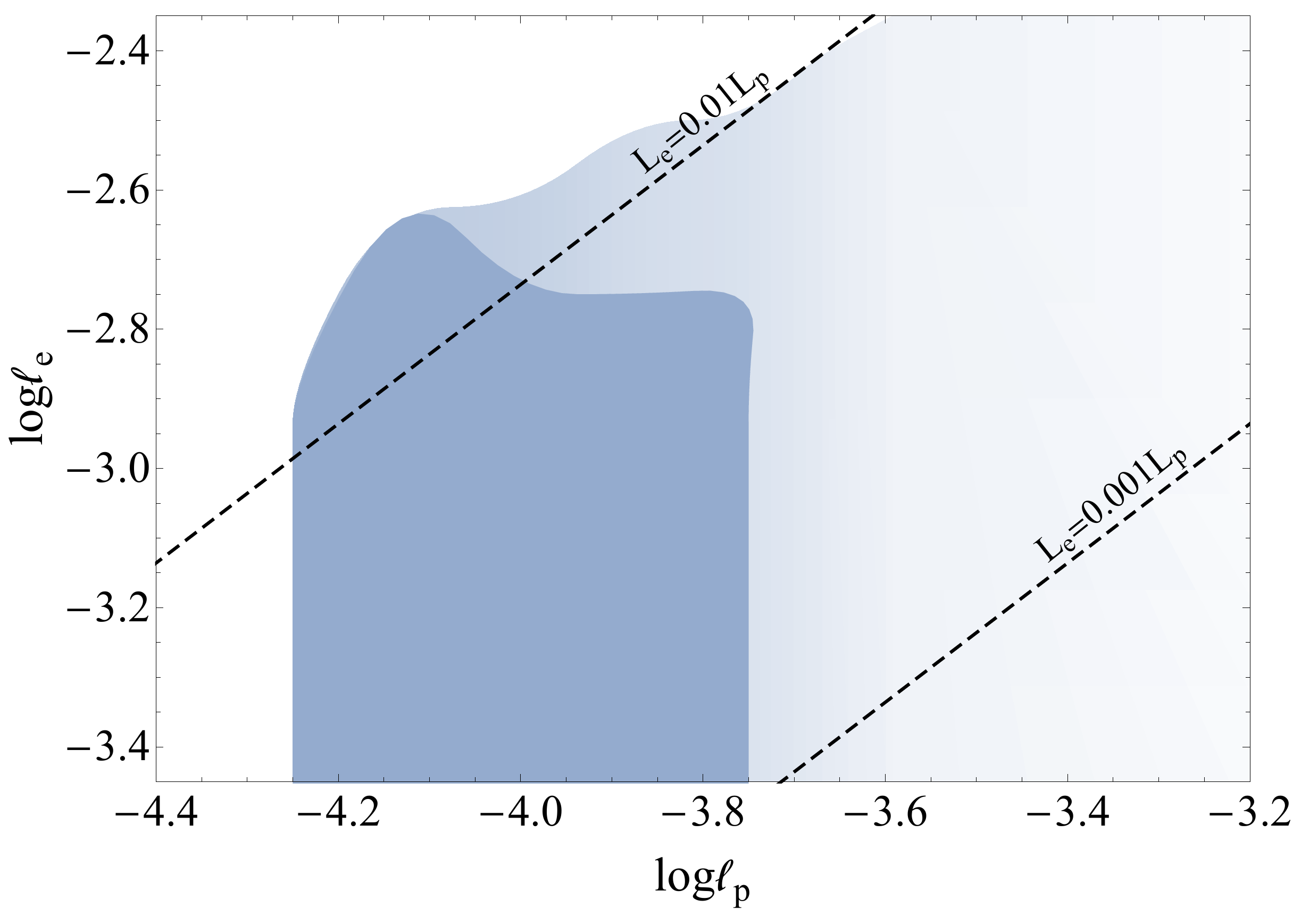} 
\caption{The parameter space $\log \ell_{\rm e}$-$\log \ell_{\rm p}$ of hadronic supercriticalities for fixed $R=10^{12}$ cm, $B=10^{3.5}$G and $\gamma_{\rm max}=10^{6.5}$. The dark coloured region corresponds to limit-cycle solutions. The colour gradient indicates the progressive transition from violent relaxations to a high compactness steady-state solution. No critical proton compactness can be defined in the white-coloured region. The dashed lines correspond to an electron-to-proton luminosity ratio of $0.1\%$ and $1\%$.}
\label{Fig:7}
\end{figure}

Fig.~\ref{Fig:7} shows the $\ell_{\rm e}-\ell_{\rm p}$ parameter space of hadronic supercriticalities for one choice of $R,B$, and $\gpmax$ values, and note that similar results are found for other sets of parameters. Supercriticality occurs for parameter combinations drawn from the grey-coloured region. Particularly, the dark-coloured region corresponds to solutions with limit-cycle behaviour, while the colour gradient indicates the progressive transition from violent relaxations to high compactness steady-state solutions. A few more things about Fig.~\ref{Fig:7} are worth mentioning.
\begin{itemize}
 \item Low values of $\ell_{\rm e}$ have absolutely no effect on the manifestation of the supercriticality, as expected.
 \item High values of $\ell_{\rm e}$ bring the system to a steady state by weakening the limit-cycle behaviour. Primary electrons also contribute to the production of photons via synchrotron and Compton processes. If $\ell_{\rm e}$ is high enough, then the resulting photon number densities will also be high, thus causing proton cooling through photohadronic processes and leading the system to a steady state.
 In this particular example, the transition to a steady-state occurs for values of $\ell_{\rm e}$ corresponding to electron luminosities $\gtrsim 0.01 L_{\rm p}$.  Moreover, the transition from a limit-cycle behaviour to a steady state is also a function of $\gamma_{\rm e,max}$, with lower values increasing the required value of $\ell_{\rm e}$ (not explicitly shown in the figure). 
\end{itemize}
It is also noteworthy that the injection of primary relativistic electrons does not decrease the required value of $U_{\rm p,crit}$. The main reason behind this is that primary electrons do not contribute to the networks leading to supercriticality, as this is an intrinsic property of a purely hadronic system. Finally, similar results are found if one considers the presence of external radiation fields (i.e., not produced by the relativistic particles in the source). In this case, external photons would serve as an extra cooling agent for the relativistic protons, and their presence at high densities would stabilize the hadronic system.

\begin{figure}
 \centering
 \includegraphics[width=\linewidth]{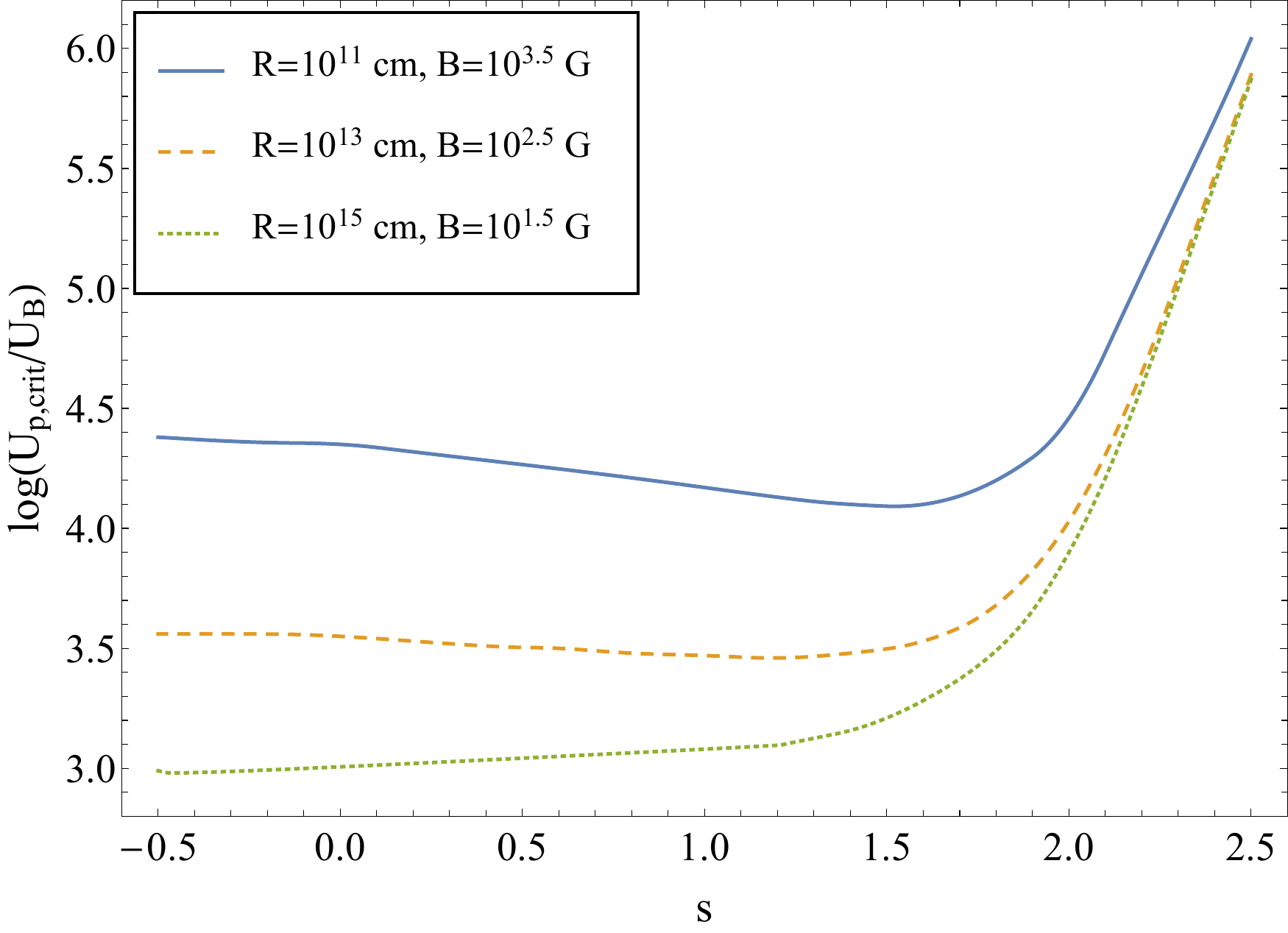} 
\caption{The dependence of the ratio $U_{\rm p,crit}/ U_{\rm B}$ (in logarithmic scale) on the power-law slope of the injected protons $s$,  for $\gamma_{\rm p,max}=10^{6.5}$ and different combinations of $B$ and $R$ (see inset legend).}
 \label{Fig:8}
\end{figure}
\subsection{Power-law index of the proton energy spectrum}\label{Sec:5:2}
Up until this section, all results have been obtained for relativistic protons with a power-law of index $s=2$. This is a standard value obtained, for example, from first order Fermi acceleration at strong non-relativistic shocks. However, the spectral index, which depends on the acceleration process and is a free parameter for the purposes of this study, can affect the onset of hadronic supercriticality.

Here, we examine the dependence of supercriticality on the slope $s$ of the the power-law proton energy distribution. Our results are shown in Fig.~\ref{Fig:8} where the ratio of the critical proton energy density to the magnetic field energy density, $U_{\rm p,crit}/U_{B}$, is given as a function of the spectral index for $\gamma_{\rm p,max}=10^{6.5}$ and three pairs of $R,B$ values (see figure's inset legend). In all cases, the dependence of $U_{\rm p,crit}/U_{B}$  on $s$ is similar:   For steep spectra (i.e., $s>2$) the ratio $U_{\rm p,crit}/U_{B}$ is high, but then decreases as the proton energy spectrum flattens, reaching a broad minimum around $s\simeq 1.5$. For even flatter spectra ($s\lesssim 1.5$), the critical proton density becomes almost independent to the slope. 

These findings can be qualitatively understood, if one recalls that only protons above a certain energy (i.e., $\gamma_{\rm p} > \gamma_{\rm p,crit}$) are relevant for the onset of supercriticality through the relevant energy threshold conditions (see Section~\ref{Sec:4} and Fig.~\ref{Fig:5}).
In other words, $\gamma_{\rm p,crit}$ is effectively the minimum Lorentz factor for computing the critical energy density of the proton distribution. 
As in our calculations we have assumed that $\gamma_{\rm p,min}=1$, steep spectral indices mean that a large fraction of protons have $\gamma_{\rm p,min} < \gamma_{\rm p} < \gamma_{\rm p,crit}$, thus contributing to the total energy density but without being relevant for supercriticality. Thus, a higher total proton energy density is needed to compensate for the large number of ``inactive'' protons. For flat indices (i.e., $s\lesssim 1.5$), we find a very weak dependence of $U_{\rm p,crit}/U_{B}$ on $s$, as the proton critical energy density is now mainly determined by the maximum Lorentz factor of the distribution, which is kept fixed.

\begin{figure}
 \centering
  \includegraphics[width=\linewidth]{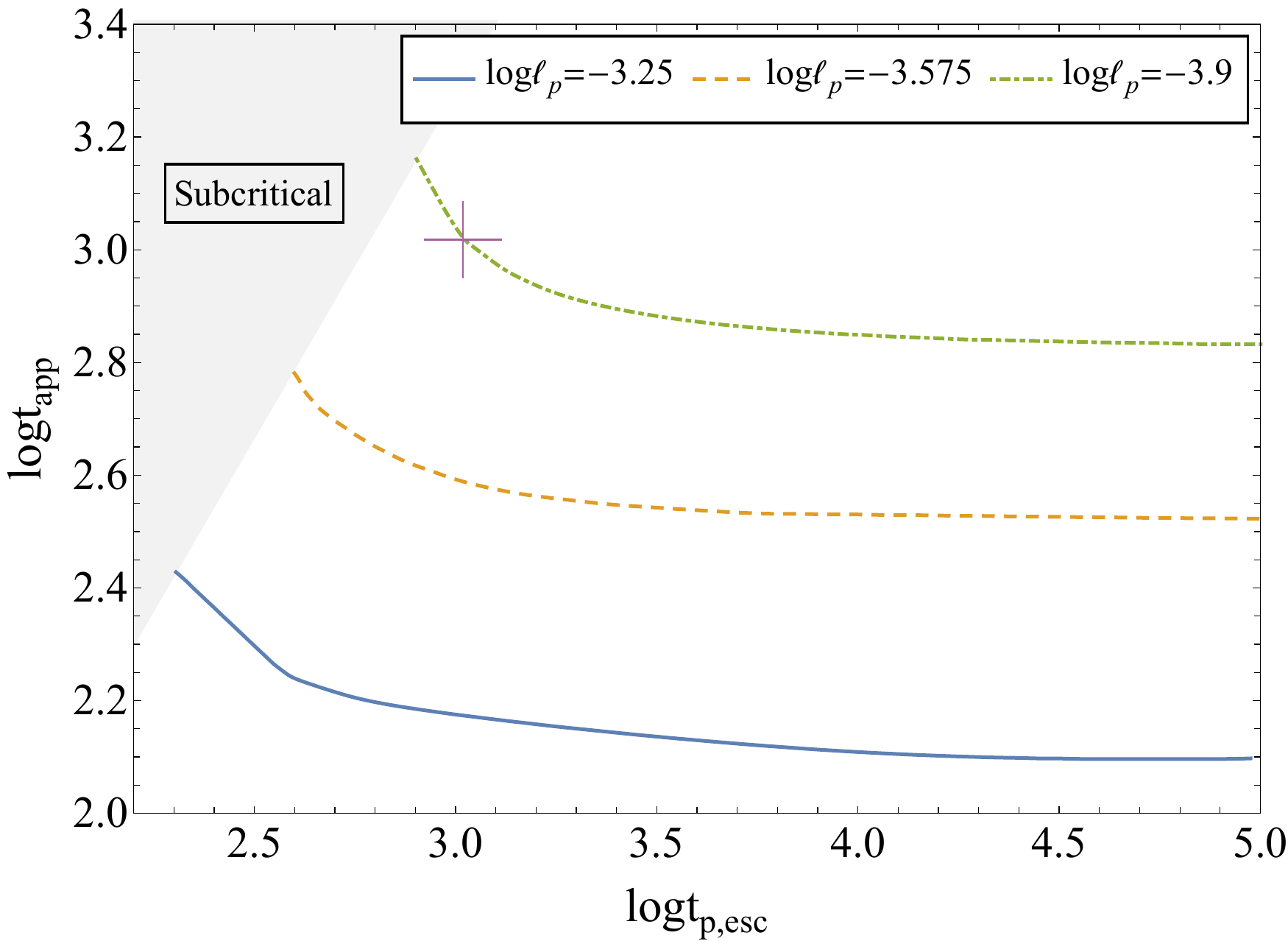} 
\caption{The appearance time of limit cycles as a function of the escape timescale of the injected protons (in logarithmic scale). Both timescales are in units of $t_{\rm cr}$. We show results for $R=10^{13}$~cm, $B=10^{2.75}$~G, $\gamma_{\rm p,max}=10^{6}$, and different values of $\lp$ (listed in the inset legend).
The cross depicts the point where $t_{\rm p,esc}=t_{\rm app}=10^3 \, t_{\rm cr}$, while in the gray area the system is subcritical.}
\label{Figtapp}
\end{figure}
\subsection{Proton escape timescale }
The proton escape timescale $t_{\rm p,esc}$ is another parameter of the problem that was kept fixed ($t_{\rm p, esc}=10^3\tcr$) for the purposes of our previous analysis. It is also a parameter which controls $\ell_{\rm p,crit}$, i.e., the value of the proton injection compactness parameter which is required to bring the hadronic system into supercriticality, through the solution of the equation $t_{\rm app}(\ell_{\rm p, crit})=t_{\rm p, esc}$. Note, however, that the required critical proton density is independent of the choice of $t_{\rm p,esc}$ as, in the absence of strong losses, one can write $n_{\rm p,crit}\propto \ell_{\rm p, crit}\, t_{\rm p,esc}$. Clearly, when $t_{\rm p,esc}$ is small, implying a fast escape timescale, one would require a high value of $\ell_{\rm p,crit}$ in order to meet the marginal stability criterion and vice versa.

Fig.~\ref{Figtapp} depicts the behaviour of $t_{\rm app}$ as a function of $t_{\rm p,esc}$ for three different values of $\ell_{\rm p}$. The top curve corresponds to $\ell_{\rm p,crit}$, and the cross denotes the point $t_{\rm app}=t_{\rm p, esc}=10^3\tcr$. By increasing $t_{\rm p,esc}$ we get an asymptotic behaviour for $t_{\rm app}$ which remains close to $10^3\tcr$ and justifies the used criterion for entrance to supercriticality. However, for given $\ell_{\rm p}$ there exists a minimum value of $t_{\rm p,esc}$ below which the marginal stability criterion cannot be met and the system becomes subcritical (shaded region in the figure). 

We find the same overall behaviour if we are to repeat the procedure for different values of $\ell_{\rm p}$. For $\ell_{\rm p} > \ell_{\rm p,crit}$, as are the two cases shown in Fig.~\ref{Figtapp} (dashed orange and solid blue lines), the asymptotic value of $t_{\rm app}$ decreases, and the same happens for the minimum value of $t_{\rm p,esc}$ required to bring the system into supercriticality.

\section{Astrophysical Implications}\label{Sec:6}
In this section, we discuss hadronic supercriticalities in the context of compact astrophysical sources. For sources that invoke relativistic motions of the emitting region (e.g., GRBs and AGN jets), all quantities, such as critical energy densities and length scales, refer to their co-moving frame values (unless stated otherwise). The application of our results to compact astrophysical sources is inviting, as hadronic supercriticalities:
\begin{itemize}
    \item produce spontaneous rapid bursts which last a few crossing times of the source. Therefore, they can in principle be applied to both transient astrophysical systems, like GRBs \citep{PM18}, and persistent yet flaring sources, like AGN cores and jets.
    \item appear on length scales relevant to GRBs ($\sim 10^{11}-10^{14}$ cm) and jetted AGN ($\sim 10^{15}-10^{17}$ cm). As the onset of supercriticalities depends on the source compactness, we find that supecriticality weakens for larger radii  (i.e., the growth rate of instability decreases) and eventually disappears; for instance, we do not find any signs of supercriticality at sizes of the radio lobes of AGN ($\gtrsim 10^{20}$~cm).
    \item appear if $\gamma_{\rm{p,max}}>10^3$. Such proton energies can be accommodated by most of the proposed models for particle acceleration and they could produce neutrinos with energies as low as a few hundreds of GeV.
 \end{itemize}
\begin{figure}
 \centering
 \includegraphics[width=\linewidth]{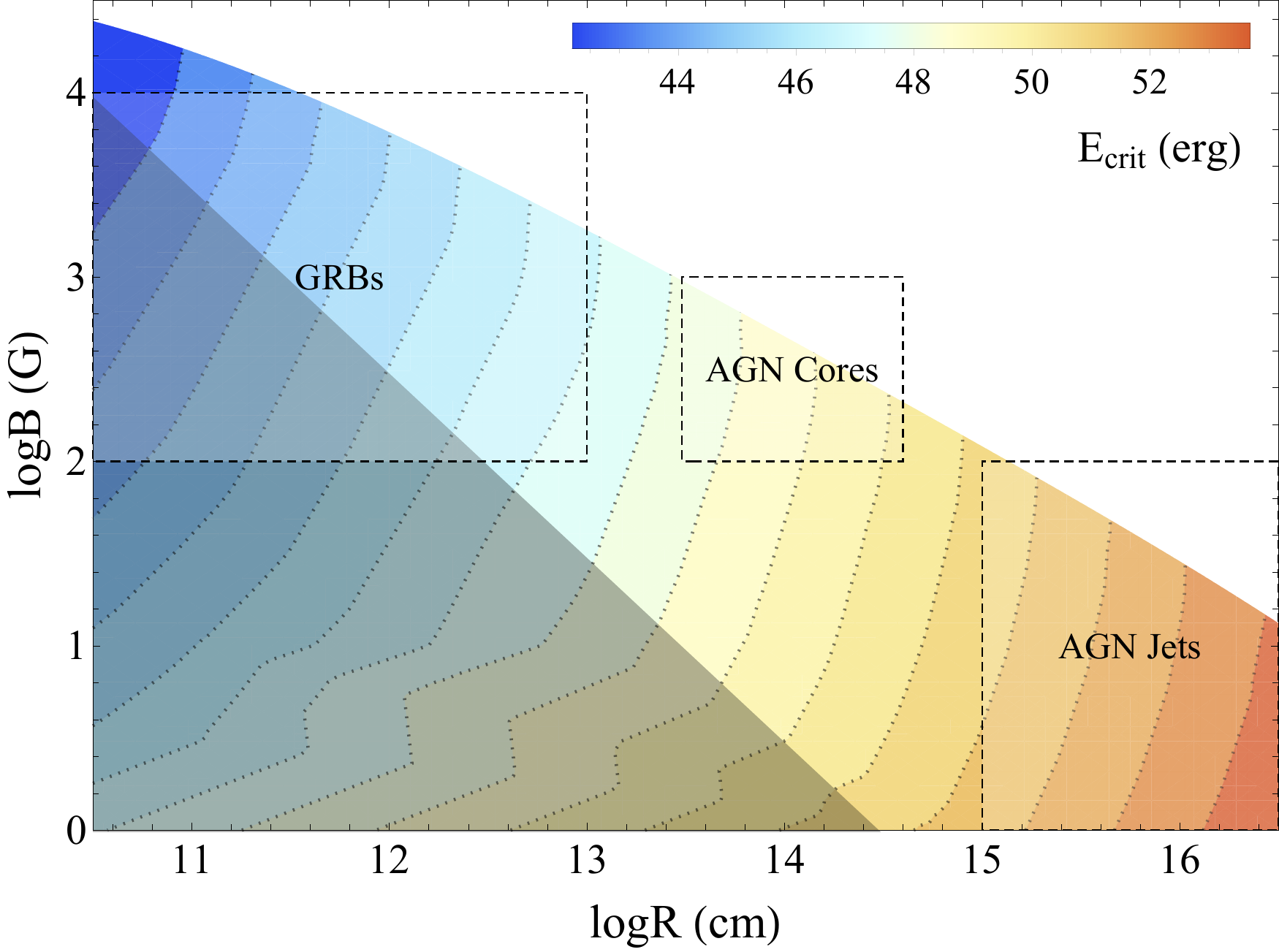}
\caption{Log-log plot of the $B-R$ phase space  for $\gamma_{\rm p,max}=10^{8}$. Contours of $E_{\rm crit}$ are overplotted (dashed lines).
The shaded area (bottom left corner) corresponds to proton gyroradii $r_{\rm g}$ larger than the source radius $R$, and is therefore forbidden. Inset legends indicate parameter ranges of high-energy   astrophysical sources. For relativistically moving sources, $B$ and $R$ refer to their co-moving frame values. }
\label{Fig:Epcrit}
\end{figure}
 We provide next some energetic estimates of supercritical bursts as related to the size and magnetic field of the source. For this purpose, we define a quantity of supercriticality which is potentially relevant for astrophysical applications, i.e., the critical proton energy, $E_{\rm crit}$. This is the amount of energy in relativistic protons that is required for a source to become supercritical, and it can be calculated numerically for any set of initial parameters that drive the system into supercriticality from the relation
\eqb
 E_{\rm crit} = 4\pi R^2 m_{\rm p} c^2  \sth^{-1} \ell_{\rm p, crit}
 \eqe
Fig.~\ref{Fig:Epcrit} shows contours of $E_{\rm crit}$ in the $B-R$ plane for $\gamma_{\rm p,max}=10^{8}$ (see inset colour bar). $E_{\rm crit}$ cannot be defined in the upper right corner of the plot which is intentionally left blank. The positions of the inset boxes indicate very broadly the range of $B$ and $R$ values relevant for GRBs, AGN cores, and jets. The minimum amount of proton energy required to produce a supercritical outburst in photons increases as we move to sources with weaker magnetic fields and larger sizes.  
\begin{figure}
 \centering
 \includegraphics[width=\linewidth]{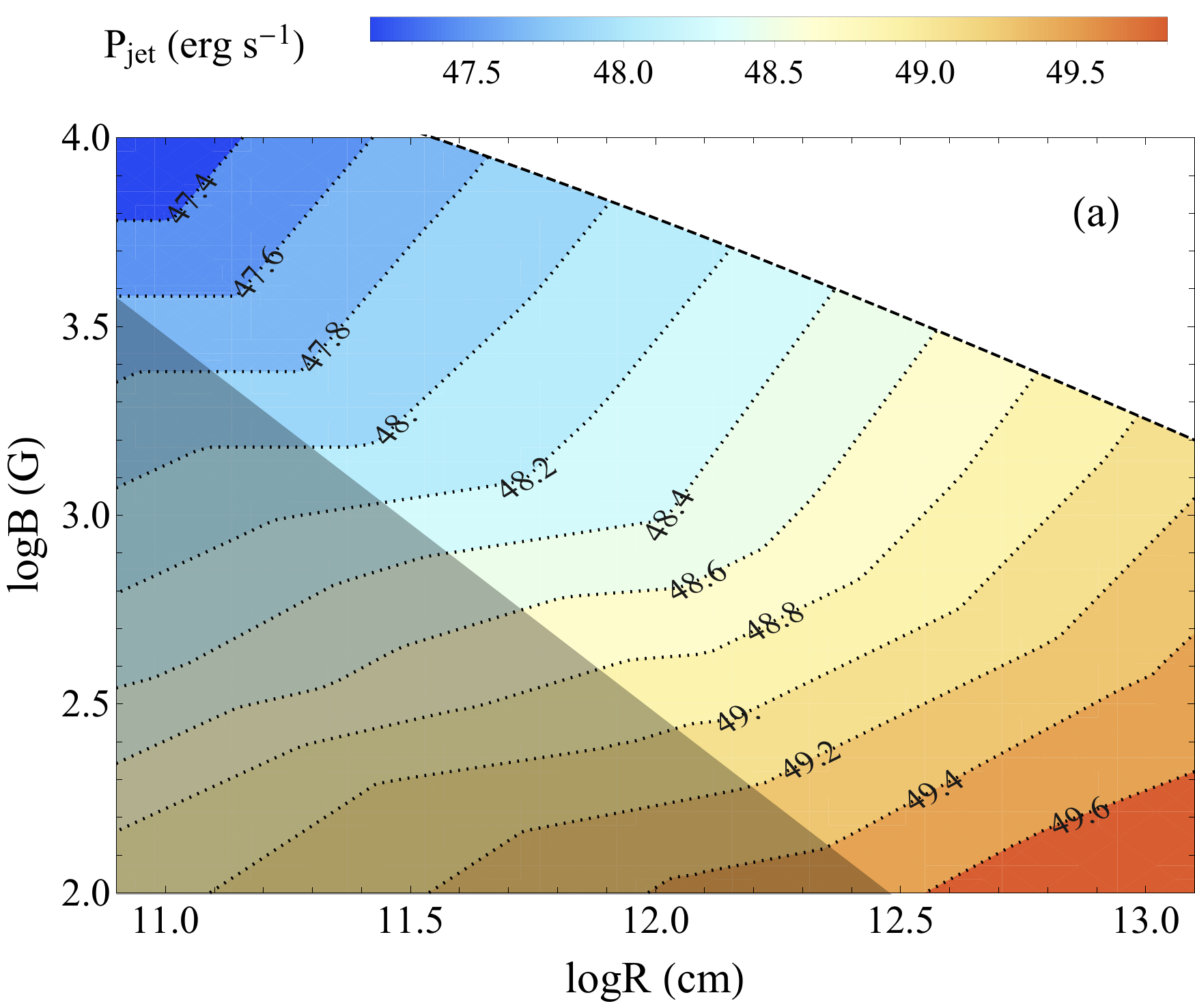}
 \includegraphics[width=\linewidth]{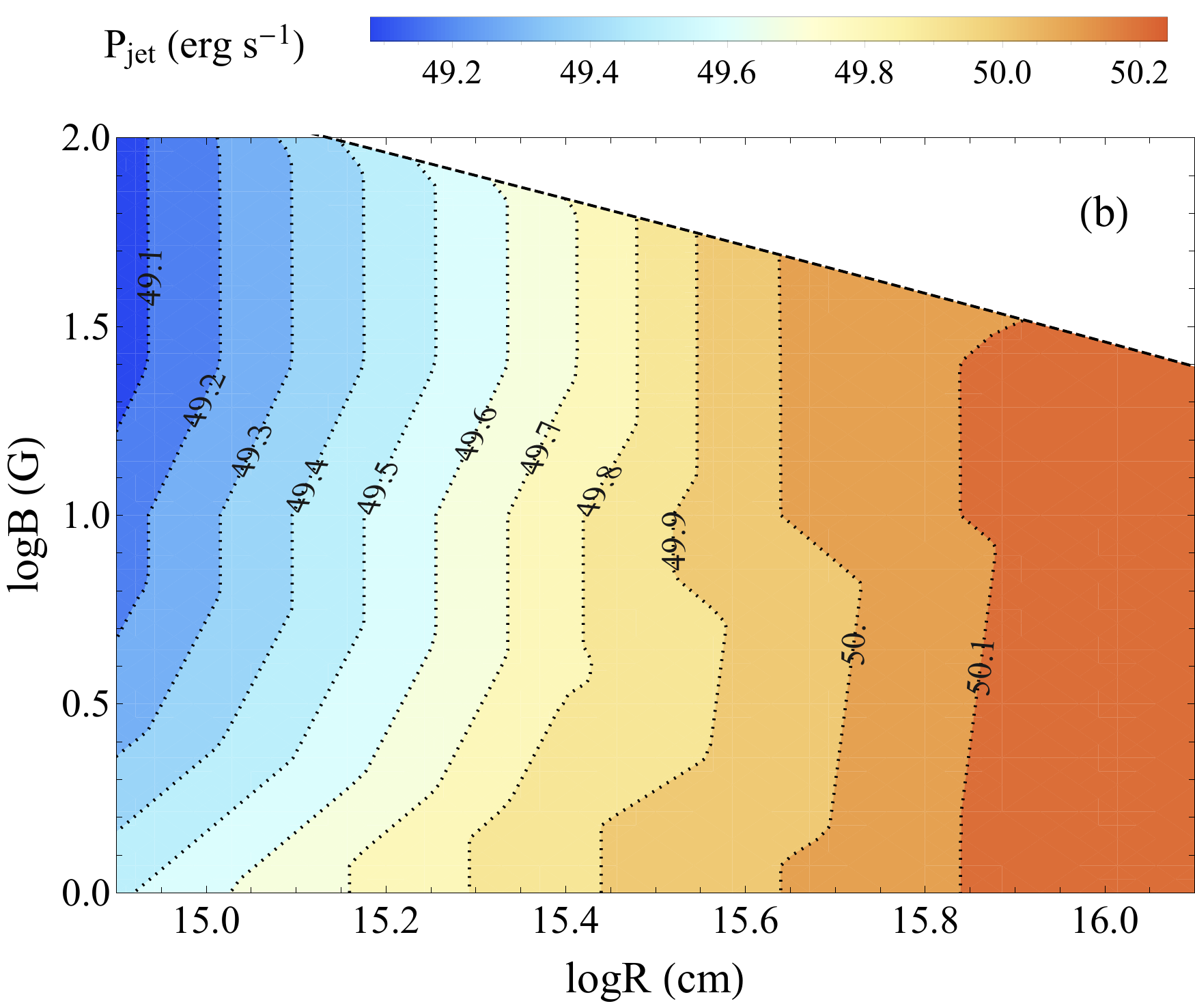}
 \caption{
Same as in Fig.~\ref{Fig:Epcrit}, but for contours of the absolute power of a two-sided relativistic magnetized jet, $P_{\rm jet}$, containing relativistic protons with the critical proton density, for GRBs (panel a) and AGN jets (panel b), assuming a bulk Lorentz factor of 200 and 20, respectively. }
\label{Fig:jetpower}
\end{figure}
The absolute power of a two-sided relativistic magnetized jet containing relativistic protons with the critical proton density can be written as \citep[e.g.,][]{PD16} 
\eqb
P_{\rm jet}\simeq \frac{8}{3}\pi R^2 c\Gamma^2 \left(U_{\rm p, crit} + U_{\rm B}\right),
\label{eq:pjet}
\eqe
where $U_{\rm p,crit}= 3 E_{\rm crit} /4 \pi R^3$. Fig. ~\ref{Fig:jetpower} shows contours of the jet power in the $B-R$ plane for $\gpmax=10^{8}$ and two values of the Lorentz factor relevant to GRBs (panel a) and AGN jets (panel b), i.e., $\Gamma=200$ and 20, respectively. As a rule of thumb, higher values of $P_{\rm jet}$ correspond to sources of larger radii. Despite the higher $\Gamma$ value of GRBs, the critical jet power is lower than the one obtained for AGN jets due to the smaller co-moving source radii related to GRBs. Although the predicted highly super-Eddington jet powers disfavour the applicability of hadronic supercriticalities to the steady AGN emission, they can still be of relevance for flaring episodes. We plan to explore this possibility in a future publication.

\begin{figure}
 \centering
\includegraphics[width=\linewidth]{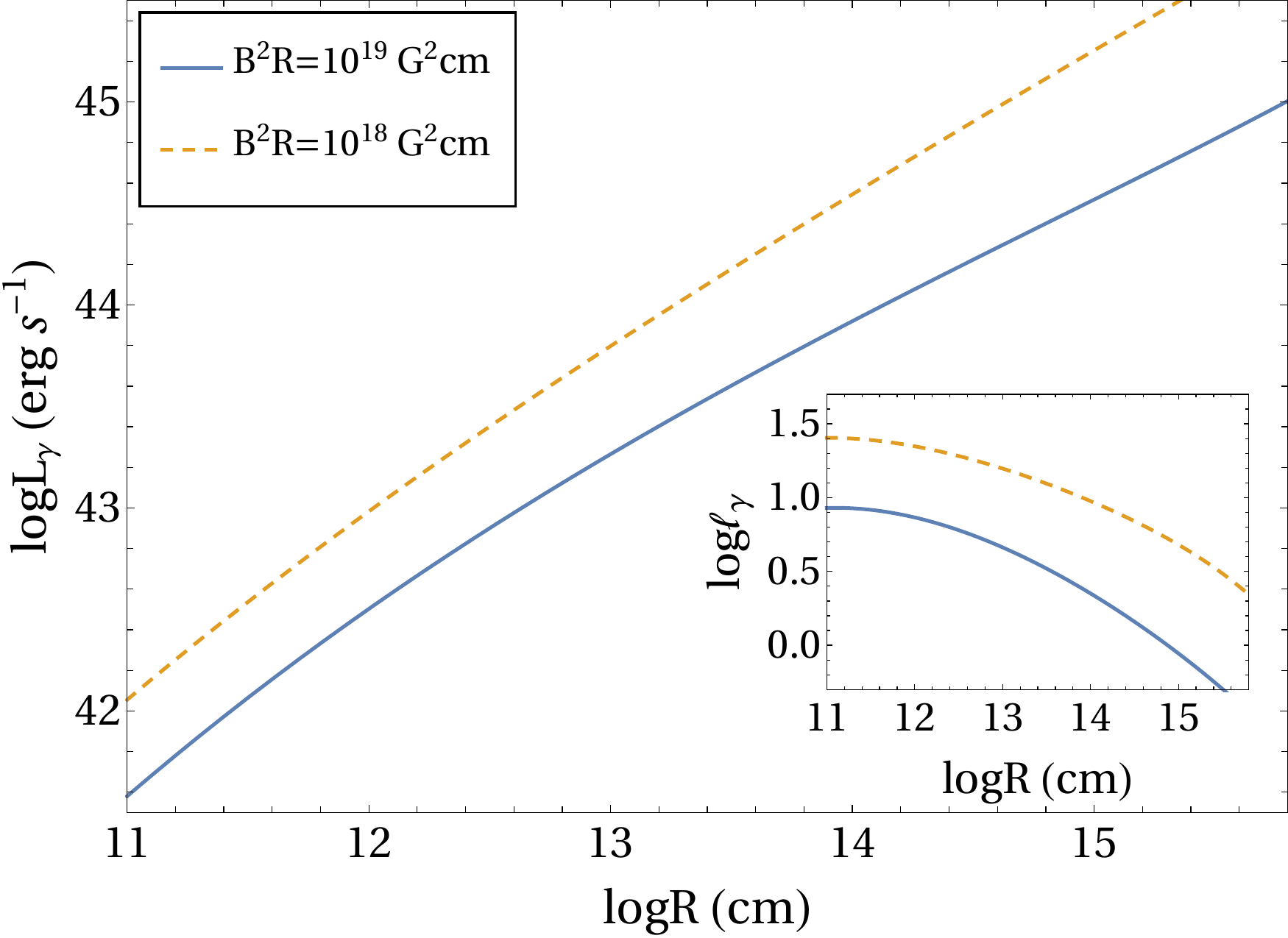}
\caption{Log-log plot of the photon luminosity at the peak of the outburst caused by hadronic supercriticality versus the size of the source $R$ for two cases: (a) $B^2R=10^{18} \rm{~G^2cm}$ (dashed orange line) and (b) $B^2R=10^{19} \rm{~G^2cm}$ (solid blue line). The inset plot shows the corresponding values of the photon compactness versus $R$. All other parameters are same as in Fig.~\ref{Fig:Epcrit}. }
 \label{Fig:LvsR}
\end{figure}

Fig.~\ref{Fig:LvsR} shows the luminosity $L_\gamma$ and the corresponding photon compactness $\ell_\gamma$ (inset)
at the peak of the outburst from the runs that produced Fig.~\ref{Fig:Epcrit} for two combinations of $B$ and $R$ that bring the system into the supercritical regime: $B^2R=10^{18} \rm G^2 \, cm$ (dashed orange line)  and $B^2R=10^{19} \rm G^2 \, cm$ (solid blue line)\footnote{The choice of $B^2R$=constant implies that the proton cooling timescale remains constant for the runs that obey this relation.}. The curves in the main plot of Fig.~\ref{Fig:LvsR} show the maximum luminosity of a supercritical flare as a function  of the source radius $R$. 
For even higher values of $B$ the system will reach a steady state and all flaring activity will cease. We note, finally, that this luminosity assumes that the source is at rest. If the source is moving relativistically with respect to the observer, one has to transform the luminosities to the observer's frame, by using the appropriate Doppler boosting. 

In general, the photon compactness is a measure of the opacity for photon-photon pair production. This can be understood as follows. Writing the $\gamma \gamma$ cross-section as ${\rm d}\sigma(\epsilon,\epsilon_\gamma)/{\rm d}\epsilon \approx (\sth/3)\epsilon \delta(\epsilon-2/\epsilon_\gamma)$ \citep{1985ApJ...294L..79Z}, where $\epsilon, \epsilon_{\gamma}$ are the low-energy and high-energy photon energies respectively (in $m_{\rm e} c^2$ units), we can estimate the optical depth as $\tau_{\gamma \gamma}(\epsilon_{\gamma}) \approx \ell_{\gamma}(2/\epsilon_\gamma) \approx (\sth R/3) \epsilon n(\epsilon)|_{\epsilon=2/\epsilon_{\gamma}}$. Although the inset plot in Fig.~\ref{Fig:LvsR} shows the photon compactness integrated over energies, $\ell_{\gamma}$, it can still be used  as a proxy for $\tau_{\gamma \gamma}$. Given that the compactness decreases with increasing radius, the optical depth for the attenuation of high-energy photons is respectively lower in larger sources.

\begin{table*}
\centering
\begin{tabular}{l c c c c }
\hline \hline
{Source} & {$B$} (G)  & {$R$} (cm) & $t_{\rm v, obs}$ (s)  & $L^{\rm pk}_{\gamma, \rm obs}$ (erg s$^{-1}$) \\  
\hline  \smallskip  
{GRBs} & $10^{2}-10^{4}$ & $10^{11}-10^{13}$ & $0.018-1.8$ & $5 \times 10^{50}-10^{53}$ \\   \smallskip
{AGN cores} &$10^{2}-10^{3}$&$3 \times 10^{13}-3 \times 10^{14}$& $10^{3}-10^{4}$ &$2 \times 10^{43}-3 \times 10^{44}$\\   \smallskip
{AGN jets} &$1-10^{2}$& $10^{15}-10^{16}$ &$1.6 \times 10^{3}-1.6 \times 10^{4}$& $5 \times 10^{49}-10^{51}$ \\
\hline 
\end{tabular}
\caption{Typical variability timescale and isotropic luminosity at the peak of the outburst caused by supercriticality (in the observer's frame) for $B$ and $R$ ranges relevant for GRBs, AGN cores and AGN jets, with  $\Gamma$ equal to 200, 1 and 20, respectively. In all cases, $\gamma_{\rm p,max}=10^{8}$.}
    \label{tab:LvsR}
\end{table*}

Finally, Table~\ref{tab:LvsR} shows typical values of the isotropic luminosity $L^{\rm pk}_{\gamma, \rm obs}$ calculated at the peak of the first outburst (in the observer's frame) for a wide range of $R, B$ values that are relevant for GRBs, and AGN (cores and jets). The table also lists the typical variability timescale, $t_{\rm v, obs}$, which is the respective light-crossing time (in the observer's frame). To compute $L^{\rm pk}_{\gamma, \rm obs}$ and $t_{\rm v, obs}$, we have made the appropriate Lorentz transformations from the source's co-moving frame to the observer's frame, assuming a Doppler factor $\mathcal{D} \approx \Gamma$, with $\Gamma=200$ for GRBs, $\Gamma=20$ for AGN jets, and $\Gamma=1$ for AGN cores. The exact values of $L_{\gamma}$ depend on the choice of $\gamma_{\rm p,max}$, which here was assumed to be $10^{8}$. Keeping in mind that a substantial fraction of the proton energy is radiated  during outbursts, Fig.~\ref{Fig:5} implies that outbursts will have roughly the same photon luminosity for $\gamma_{\rm p,max}\sim 10^6-10^8$. Although supercriticality  produces higher photon luminosities for $\gamma_{\rm p,max}<10^{6}$, this happens at the expense of higher proton energy densities.  

Even though we did not make any attempt to simulate specific sources, the predicted luminosities and variability timescales are of the same order of magnitude as those observed in GRBs \citep[for a recent review, see][]{zhang2018}. Similarly, for $R,B$ values relevant to AGN cores, we predict photon flares with peak luminosities $\sim 10^{43}-10^{44}$~erg s$^{-1}$ with hour-long to day-long variability timescales. Interestingly, TeV emission with variability on day-long timescales and luminosities $\mathcal{O}( 10^{41})$~erg s$^{-1}$ has been detected from the radio galaxy M87 \cite{Aharonian06, Aleksic12}, thus pointing to a compact $\gamma-$ray source. Although we have not discussed the high-energy photon spectrum in this work, supercritical hadronic flares in AGN cores appear a promising topic for further study. For parameters related to AGN jets, the predicted peak luminosities are much higher than typically observed for TeV blazars, even if a lower $\Gamma$ value is adopted. However, these estimates match the values reported for the most luminous $\gamma-$ray flares from flat spectrum radio quasars \citep[e.g.,][]{nalewajko2013}. Still, the properties of the resulting $\gamma-$ray spectrum need further investigation. Finally, it is interesting to note that, at least under the assumption of a constant proton injection, the model predicts limit cycles with periods of a few hundred crossing source times (see Fig.~\ref{Fig:1}). For parameter values relevant to AGN jets, the model-predicted periods can range from months to years, which is consistent with the recently reported quasi-periodic $\gamma$-ray emission of some blazars \citep{2015ApJ...813L..41A,2020Otero}.

\section{Summary and Discussion}\label{Sec:7}
Supercriticalities are an intriguing property of hadronic relativistic plasmas. They are, essentially, radiative instabilities that appear when large amounts of energy are stored in slowly cooling protons. The inability of protons to efficiently radiate their energy away can cause their energy density to exceed some critical value and initiate a runaway process which results in the explosive transfer of the proton-stored energy into electron-positron pairs and radiation. The idea has some analogies to supercritical nuclear piles in the sense that there are some clearly identifiable networks of processes (loops) that operate leading to runaway -- however, in our case it is the photons, rather than the neutrons, which are the agent. Despite the fact that the concept of hadronic supercriticalities was put forward almost three decades ago (\citealt{Stern91}, \citealt{Stern92}, KM92), there has not been as of yet a comprehensive study neither of the parameter space that leads to such situations, nor of the behaviour of the source during the onset of the non-linear outgrowth. In this work, we have addressed both questions with more emphasis on the former, thus  providing a complete roadmap to hadronic supercriticalities. 

The runaway that ensues when protons enter the supercritical regime produces, in most cases, a flare in photons, and the obtained light curves are distinctively different from the subcritical ones leading to steady-state -- see Fig. \ref{Fig:1}. The subsequent behaviour of the system in supercriticality depends on various factors. If the protons continue to be injected in the system at the same rate -- an assumption that we made in our analysis -- then the system can exhibit either a (quasi-periodic) limit-cycle behaviour, or a violent relaxation ending up to a steady state. The limit-cycle behaviour can be explained on physical grounds from the fact that supercriticality is caused when protons of a certain energy reach a certain critical density -- see KM92. The exponentiation of photons causes strong proton losses through photohadronic interactions. Therefore, protons are pushed back to the subcritical regime and the runaway dies out. If, however, the injection of fresh protons continues, then once again protons will accumulate, and eventually they will enter the supercritical regime causing a new photon runaway and the cycle repeats. This is the description of a classic ``prey-predator'' system, as it was first shown in PM12 (see also  Appendix~\ref{App:B}). 

In order to set the framework that would enable us to examine the supercriticality, we have used the premises of the one-zone model that has been widely used both for leptonic and hadronic modelling of high-energy sources. Protons were assumed to be injected with a constant rate inside a spherical source of radius $R$ and magnetic field $B$ while, at the same time, they were allowed to cool by synchrotron radiation and photohadronic processes and to physically escape from the source. 
The evolution of the system was followed by means of a numerical code that solves the continuity equations for protons, electrons, and photons. Assuming a generic power-law energy spectrum for the injected protons and using the high-energy cutoff $\gamma_{\rm p,max}$ and the injection compactness $\ell_{\rm p}$ (equation \ref{eq:qp}) as the main parameters, we performed runs for various combinations of $R$, $B$, and $\gamma_{\rm p,max}$. Having as guidelines previous analytical works (KM92, PM12) and parameters of astrophysical interest, we chose characteristic values of $R$ ranging from $10^{11}$~cm to $10^{16}$~cm, that are relevant to GRBs and AGN (cores and jets), and let the  magnetic field strength vary from 1~G to 100~kG. We also considered a very wide range of $\gamma_{\rm p,max}$ values extending from $10^3$ to $10^9$ (see  Table \ref{Table:1}).

We demonstrated that (for most parameters) there exists a critical value of the injected compactness, $\ell_{\rm p,crit}$, that drives the system into a state characterised by rapid photon outgrowth, even if initially the source did not contain any photons -- see e.g., Fig.~\ref{Fig:2}. We were able to determine unequivocally the value of $\ell_{\rm p,crit}$, because, in the vast majority of the cases, the system enters very abruptly in the supercritical regime, i.e., for $\ell_{\rm p}=\ell_{\rm p,crit} -\delta\ell_{\rm p}$ the system is still stable and it suffices a small increment of $\delta\ell_{\rm p}\ll \ell_{\rm p}$ to push it into the supercritical regime, as illustrated in Fig.~\ref{Fig:1}. The phase space of hadronic supercriticalities and the temporal behavior of the system in the non-linear phase of the instability outgrowth are summarized in Fig.~\ref{Fig:3}.

In the parameter exploration, we assumed that pions and muons decay instantaneously to their secondary particles. However, in the presence of strong magnetic fields and high proton energies, pion and/or muon synchrotron cooling becomes relevant, and may affect the photon (and neutrino) spectrum. Pions and muons with Lorentz factors $\gamma_\pi$ and $\gamma_\mu$, respectively, can cool due to synchrotron radiation before they decay, if the following conditions are satisfied, $\gamma_{\pi} B > 8 \times 10^{11}$~G and $\gamma_{\mu} B > 6 \times 10^{10}$~G \citep[see also equation 14 in][]{PGD14}.  Assuming energy equipartition between secondaries of the decaying particle, and an inelasticity of $\kappa_{\rm p}\simeq 0.25$ , one readily finds that $\gamma_{\pi} = \kappa_{\rm p} (m_{\rm p}/m_{\pi}) \gamma_{\rm p}\simeq 0.7\gamma_{\rm p}$, and $\gamma_{\mu} = (m_{\pi}/3 m_{\mu}) \gamma_{\pi} \simeq 0.4 \gamma_{\pi}$, where $m_{\pi}$ and $m_{\mu}$ are the pion and muon rest masses, respectively. Thus, pion and muon synchrotron cooling becomes relevant for $\gamma_{\rm p,max}=10^8$ and $B \gtrsim 600$~G (see panel d in Fig.~\ref{Fig:3}). Even in this regime, however, we do not expect  major modifications of our results mainly for two reasons. First, pions and muons will produce additional synchrotron photons with typical energies $b\gamma^2_{\pi} (m_e/m_\pi)$ and $b \gamma^2_{\mu} (m_e/m_{\mu})$ at the expense of secondary electrons-positron pairs. Second, neutral pions, being unaffected by radiative cooling, will continue to produce $\gamma$-rays at an energy that depends solely on the parent proton energy. Thus, their contribution to the feedback loops sustaining supercriticality will remain unaltered (for details, see Appendix~\ref{App:A}). The inclusion of synchrotron pion and muon cooling will therefore have only a small impact on the value of $U_{\rm p,crit}$ (of a factor of $\sim2$) for part of the parameter space shown in panel d of Fig.~\ref{Fig:3}.

Although hadronic supercriticalities emerge for a wide range of source parameter combinations, we showed that  all cases require a particle dominated source with $U_{\rm p,crit}\gg U_{\rm B}$. The ratio, however, decreases with increasing $B$, and can be as low as $10^2-10^3$ for strong magnetic fields (i.e., $B\sim 10^3-10^4$~G) -- see Fig. \ref{Fig:4}. Lower values of the energy density ratio can also be achieved by considering harder injection proton spectra, with slopes $s<2$  -- see Fig.~\ref{Fig:8}. We have discussed the properties of hadronic supercriticalities under the assumption of continuous injection of relativistic protons in the source, and showed that whenever their energy density (which is a function of time) exceeds $U_{\rm  pr,crit}$, the system becomes supercritical. Thus, $U_{\rm  p}/U_{\rm B} \gtrsim U_{\rm  p, crit}/U_{\rm B} \gg 1$ has to be satisfied only for a short duration of time, of the order of a few crossing times.
Thus, the necessary departure from equipartition in the source has to be viewed in light of its transient nature.

Supercriticalities seem to be an intrinsic property of hadronic systems. Although purely leptonic systems (i.e., sources with an energetically sub-dominant or even absent population of relativistic protons) can exhibit non-linear behaviour because of synchrotron self-Compton processes, they do not show any features of supercriticality due to the fast electron cooling \citep[e.g.,][]{PPM15}. Still, the presence of a population of accelerated electrons in a hadronic system does not change the key features of hadronic supercriticalites presented in this work, as long as their luminosity is lower than $0.01L_{\rm p}$ -- see Fig.~\ref{Fig:7}. Above this value, the radiation produced by electrons quenches the relevant hadronic supercriticalities. Generally, we can say that strong photon fields either external or internal to the source, e.g., from primary electrons, can lead the system quickly to a steady state as the supercriticalities under study depend on the internally produced radiation. Any other field, as the ones produced from primary electrons, acts competitively and tends to stabilize the system.

We also compared our results with those presented in the analytical studies of KM92 and PM12, and found good agreement. Particularly, we showed that the network of processes examined by KM92 and PM12 are the most important and can explain, in a very satisfactory manner, the behaviour of the system as a function of $\gamma_{\rm p,max}$ -- see Figs. \ref{Fig:4} and \ref{Fig:5}. In short, photohadronic interactions and electron synchrotron radiation are responsible for the loops at low to mid values of $\gamma_{\rm p,max}$ (KM92), while at larger values photo-quenching and the ensuing electromagnetic cascades that cool explosively the relativistic protons take over (PM12). Other loops of processes (e.g., those with inverse Compton scattering replacing (electron) synchrotron radiation) could also operate, but they do so at the cost of a higher critical proton density.

In this paper, we explored hadronic supercriticalities by injecting  pre-accelerated protons in a fixed volume, and then let particles evolve under the presence of energy losses alone. 
Alternatively, these radiative instabilities can be studied in the case where proton acceleration and radiation take place in the same region. In this scenario, protons should be accelerated to high enough energies and reach high enough densities as to satisfy both the required feedback and critical stability criteria of hadronic supercriticalities -- see Section \ref{Sec:4}. It is possible that the system will exhibit different temporal behaviour than the one studied here, especially when the saturation proton Lorentz factor (i.e., where the energy gains from acceleration are balanced by the energy losses from radiation and inelastic collisions) is close to the critical proton Lorentz factor of $P \pi S$ loop.

We have studied the properties of hadronic supercriticalities under the assumption of particle injection in a source with a fixed volume. It is, however, possible that the source itself is expanding.
In this case, the build-up of a critical proton density in the source may be inhibited by its increasing volume and/or the ensuing adiabatic energy losses. Therefore, the question that naturally arises is whether supercriticalities can even exist in such a scenario, and if yes, in what way they are manifested. Some preliminary results  which show that supercriticalities do exist even in expanding sources are presented in \cite{FM}.

\section{Conclusions}\label{Sec:8}
In this work, we provide a roadmap to hadronic supercriticalities,  i.e., radiative instabilities that can develop in hadronic relativistic magnetized plasmas. By performing a numerical investigation over a wide range of source parameters and maximum proton energies, we have determined the conditions that drive a hadronic system to supercriticality, and systematized its rich temporal behaviour during the non-linear phase of the instability growth. By isolating certain physical processes, we have identified specific networks (originally predicted by analytical calculations) that are responsible for the onset and sustainability of these instabilities. Even though hadronic supercriticalities can emerge for a wide range of astrophysically relevant parameters, a quantitative application of these ideas to high-energy compact sources, such as GRBs, AGN jets and cores, remains to be tested.  

\section*{Acknowledgements}
The authors thank the anonymous referee for constructive comments. I.F. and S.B. acknowledge that this research is co-financed by Greece and the European Union (European Social Fund-ESF) through the Operational Programme ``Human Resources Development, Education and Lifelong Learning'' in the context of the project ``Strengthening Human Resources Research Potential via Doctorate Research'' (MIS-5000432), implemented by the State Scholarships
Foundation (IKY). M.P. acknowledges support from the Lyman Jr.~Spitzer Postdoctoral Fellowship.
Correspondence regarding this work should be sent to A.~Mastichiadis (amastich@phys.uoa.gr) or M.~Petropoulou (m.petropoulou@astro.princeton.edu).





\bibliographystyle{mnras}
\bibliography{bibliography} 



\appendix 

\section[Analogies to Dynamical Systems]{A toy model of hadronic supercriticalities}\label{App:B}

In this Appendix, we show that the behaviour of the hadronic system, as described by equation 
\ref{eq:ke} and depicted in Figs. \ref{Fig:1} and \ref{Fig:2}, can be explained in simple terms with the help of a toy model.

Following KM92, we assume that a source contains monoenergetic protons of Lorentz factor $\gamma_{\rm p}$ which, when unstable, produce through the {\sl PeS} or {\sl P$\pi$S} loops (see Section \ref{Sec:4}) electron-positron pairs of Lorentz factor $\gamma_{\rm e}$. These pairs (for simplicity, we will refer to them as electrons) radiate monochromatic synchrotron photons with energy $x_{\rm s}=b\gamma_{\rm e}^2$ (in units of $m_{\rm e}c^2$), where $b=B/B_{\rm crit}$. However, at high photon luminosities, electrons can lose energy through inverse Compton losses, producing photons at energies $x_{\rm c}=\gamma^2_{\rm e} x_{\rm s}=b\gamma_{\rm e}^4$, and this acts competitively to the synchrotron losses. 

Based on the above, we consider the following four particle distributions inside the source, namely the protons having a number density $n_{\rm p}$, the electrons $n_{\rm e}$, the soft synchrotron photons with $n_{\rm s}$, and the hard Compton ones with $n_{\rm h}$\footnote{Actually there is a fifth component: soft photons produced from proton synchrotron radiation. However, their number density is assumed to be negligible, $n_{\rm s}>>n_{\rm p,s}$. }. We can then write a simplified equation for the evolution of the proton number density
\begin{equation}
\dot{n}_{\rm p} +\frac{n_{\rm p}}{\tau_{\rm p,esc}} 
= Q_{\rm p} + \frac{n_{\rm p}}{\tau_{\rm p\gamma}}
+ \frac{n_{\rm p}}{\tau_{\rm p,syn}}
\label{eq:B1}
\end{equation}
where $Q_{\rm p}$ is the injection rate and $\tau_{\rm p,esc}$, $\tau_{\rm p\gamma}$, and $\tau_{\rm p,syn}$ are the proton escape, the photohadronic and the synchrotron loss timescale, respectively (in units of $R/c$). All number densities are normalised to $\sigma_{\rm T}R$, where $\sigma_{\rm T}$ is the Thomson cross section and $R$ is the radius of the source. 

The photohadronic loss timescale can be written as 
\begin{equation}
\tau_{\rm p\gamma}= (a\sigma_{\rm{p\gamma}} n_{\rm p}(n_{\rm s}+n_{\rm h}))^{-1}
\label{eq:B2}
\end{equation}
where $a$ is the proton inelasticity, i.e., the fraction of proton energy that goes to secondaries and $\sigma_{\rm p\gamma}$ is the relevant cross-section normalised to $\sigma_{\rm T}$.

The synchrotron loss timescale can be written under these normalisations as
\begin{equation}
\tau_{\rm p,syn}=\left(\frac{4}{3}\left(\frac{m_{\rm e}}{m_{\rm p}}\right)^3\ell_{\rm B}n_{\rm p}\gamma_{\rm p}\right)^{-1}
\label{eq:B3}
\end{equation}
where $\ell_{\rm B}$ is defined by relation \ref{eq:lb}.

In analogy to  equation~\ref{eq:B1}, we can write a continuity equation for secondary electrons produced in the source
\begin{equation}
\dot{n}_{\rm e} +\frac{n_{\rm e}}{\tau_{\rm e,esc}} 
= Q_{\rm e} 
+ \frac{n_{\rm e}}{\tau_{\rm e,syn}}
+ \frac{n_{\rm e}}{\tau_{\rm e,ics}}
\label{eq:B4}
\end{equation}
where $\tau_{\rm e,esc}$ is the electron escape timescale, $\tau_{\rm e,syn}$ is the synchrotron loss timescale at (dimensionless) energy $\gamma_{\rm e}$ given by 
\begin{equation}
\tau_{\rm e,syn}=\left(\frac{4}{3}\ell_{\rm B}n_{\rm e}\gamma_{\rm e}\right)^{-1}
\label{eq:B5}
\end{equation}
and $\tau_{\rm e,ics}$ is the corresponding inverse Compton loss timescale (in the Thomson regime) given by 
\begin{equation}
\tau_{\rm e,ics}=\left(\frac{4}{3} x_{\rm s}n_{\rm e}n_{\rm s}\gamma_{\rm e}\right)^{-1}.
\label{eq:B6}
\end{equation}

In our toy model, electrons (and positrons) are produced via photohadronic processes with a typical Lorentz factor $\gamma_{\rm e}$. Thus, the electron injection term $Q_{\rm e}$ takes the form 
\begin{equation}
Q_{\rm e}=a\frac{\gamma_{\rm p}m_{\rm p}}{\gamma_{\rm e}m_{\rm e}}n_{\rm p}(n_{\rm s}+n_{\rm h})H(n_{\rm p}-n_{\rm p,crit})
\label{eq:B7}
\end{equation}
where $H(x)$ is the Heaviside function. This term simulates the fact that secondary injection starts when the proton density exceeds the critical value $n_{\rm p,crit}$. Note that the electron injection term is written so that it conserves energy, i.e., the energy lost by protons is injected as energy in the secondaries in the electron equation.

Finally, we complete the system by writing two equations for the two photon populations (soft and hard), assuming that they do not interact through $\gamma\gamma$ absorption. These equations describe the energy channeled into soft and hard photons by electron synchrotron and inverse Compton scattering, respectively.  The soft-photon equation reads
\begin{equation}
\dot{n}_{\rm s} +n_{\rm s} 
= \frac{4}{3}\ell_{\rm B}b^{-1}n_{\rm e}
\label{eq:B8}
\end{equation}
and the equation for hard photons is
\begin{equation}
\dot{n}_{\rm h} +n_{\rm h} 
= \frac{4}{3}x_{\rm s}n_{\rm s} n_{\rm e}.
\label{eq:B9}
\end{equation}
Note that, as written, the system of equations \ref{eq:B1}, \ref{eq:B4}, \ref{eq:B8}, and \ref{eq:B9}, implies that only the soft synchrotron photons $n_{\rm s}$ produce hard photons through inverse Compton scattering.
 
Assuming that $n_{\rm p}=n_{\rm e}=n_{\rm s}=n_{\rm h}=0$ initially, we can distinguish the following cases in the regime of weak magnetic fields and negligible proton synchrotron losses:
\begin{itemize} 
    \item If $Q_{\rm p}$ is such as $n_{\rm p}(\tau)<n_{\rm p,crit}$ at all times, the system never becomes supercritical. Instead, it reaches a steady state characterized by $n_{\rm p}=Q_{\rm p}\tau_{\rm p,esc}$. In this case, no secondary particles are produced, i.e., $n_{\rm e}=n_{\rm s}=n_{\rm h}=0$ at all times. This regime corresponds to the low-efficiency steady states found numerically in our study -- see e.g., Figs.~\ref{Fig:1} and \ref{Fig:2}; the only  difference with the toy model's prediction is that the photon number density is  not equal to zero, but are produced by the very inefficient proton synchrotron process (which is neglected in our toy model). 
    \item If $Q_{\rm p}$ is such as $n_{\rm p}(\tau_{*})>n_{\rm p,crit}$, the system becomes supercritical and secondary particles start being produced.  Electrons are injected by photohadronic processes (eq. \ref{eq:B4}), producing soft photons through synchrotron radiation (eq.~\ref{eq:B8}) and hard photons through Compton scattering (eq. \ref{eq:B9}). These photons, in turn, cool the protons via photohadronic processes (eq. \ref{eq:B1}). We can further distinguish the following cases:
    \begin{itemize}
        \item If ICS losses are ignored (or, equivalently, if $\tau_{\rm e,syn} << \tau_{\rm e,ics}$, the system of equations 
        is equivalent to the predator-prey equations (or Lotka-Volterra equations), and yields limit-cycle solutions, as shown in Fig.~{\ref{Fig:1}}.
        \item If the parameters are such that $\ell_{\rm s}=x_{\rm s}n_{\rm s}>>\ell_{\rm B}$ during some phase of the system's evolution, then Compton cooling becomes faster than synchrotron cooling (i.e., $\tau_{\rm e,ics}<\tau_{\rm e,syn}$), and electron energy is radiated mostly through Compton channel, producing hard photons. Thus, it does not contribute to the proton feedback loop, but  acts instead as a damping factor -- see also   PM12. 
    \end{itemize}
\end{itemize}
By numerically solving the toy model's set of equations, one can gain insight to the temporal
behaviours depicted in Fig.~ \ref{Fig:2}. The low-efficiency steady states found for parameters from the lower left corner of this figure correspond to cases with $n_{\rm p}<n_{\rm p,crit}$ at all times. When $\ell_{\rm B}$ is small and the system is in the supercritical regime (i.e., see right-hand side of the plot), then during the first outburst $\ell_\gamma \gg \ell_{\rm B}$. This drives the system to a high-efficiency steady state through a violent relaxation because of the strong Compton damping. As $\ell_{\rm B}$ increases, electron synchrotron balances inverse Compton, and the system gets into the limit cycle regime with negligible or very low Compton damping. Finally, when $\ell_{\rm B}$ increases further, the proton synchrotron term in equation \ref{eq:B1} becomes dominant, cooling the protons and never allowing them to reach the supercritical value $n_{\rm p,crit}$. In this case, the feedback loop will not operate.

\section[The gamma-Q loop]{The $\gamma Q$ loop}\label{App:A}

While the $PeS$ and $P\pi S$ loops (see Section \ref{Subsec:4.1}) are rather straightforward as they involve protons directly, the $\gamma Q$ loop (Section \ref{Subsec:4.2}) is more subtle and requires special attention. Here, we will repeat and expand on some arguments first presented in PM12. 
\begin{figure}
 \centering
 \includegraphics[width=\linewidth]{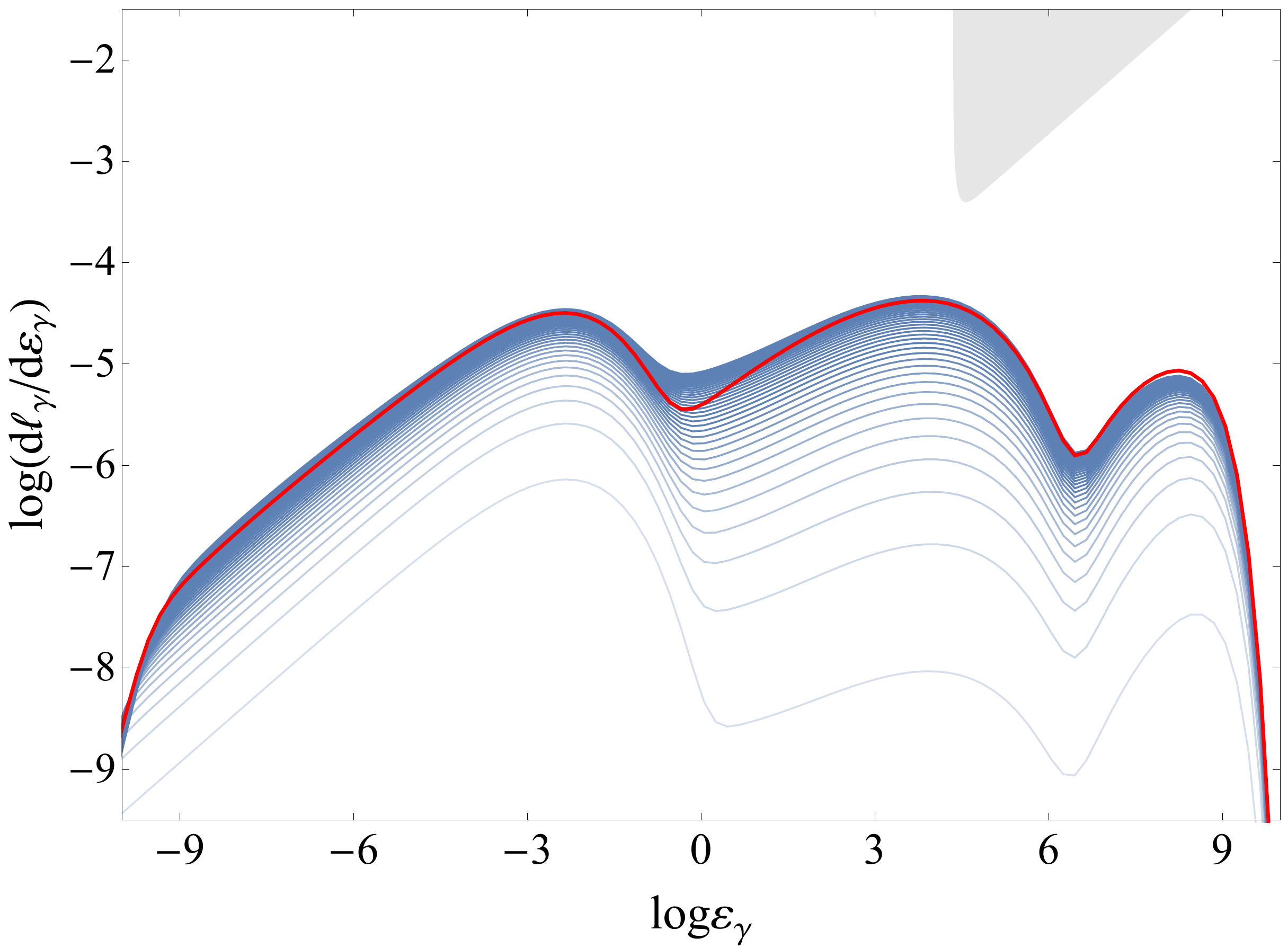} 
\caption{Snapshots, taken every 50$t_{\rm cr}$, of the multiwavelength photon spectrum as the source starts without particles and photons and eventually a steady-state is established. The red line shows the corresponding steady-state when, from all processes, only electron and proton synchrotron radiation, photopion, and $\gamma\gamma$ absorption are used. The parameters used here are: $R=10^{15}$~cm, $B=10^{1.5}G$, $\gamma_{\rm p,max}=10^{6.5}$, $\ell_{\rm p}=10^{-4.5}$.}
 \label{Fig:A1}
\end{figure}
\subsection{An example of subcritical photon emission}
Fig.~\ref{Fig:A1} shows the temporal evolution of a typical multiwavelength (MW) spectrum obtained when a proton power-law is injected in a source of radius $R$ and magnetic field $B$.  Snapshots are displayed every $50 t_{\rm cr}$ (thin blue lines), and are computed when all processes taken into account. The steady-state spectrum (computed when electron and proton synchrotron radiation, photopion production, and $\gamma\gamma$ pair production are included) is also shown (thick red line) for comparison. For the parameters used (i.e., $R=10^{15}$~cm, $B=10^{1.5}$~G, $\gamma_{\rm p,max}=10^{6.5}$, and $\ell_{\rm p}=10^{-4.5}$, similar to those used in Fig.~\ref{Fig:1}), the system is in the subcritical regime. The MW photon spectrum at all times consists of three distinct peaks. Before discussing the temporal evolution of the photon emission, let us take a closer look at the processes producing the three components of the photon spectrum.

The low-energy peak is produced by proton synchrotron radiation which peaks approximately at an energy (in $m_{\rm e} c^2$ units)
\eqb
x_{\rm p,syn}\simeq \frac{m_{\rm e}}{m_{\rm p}} b\gamma_{\rm p,max}^2
\label{eq:A1}
\eqe
The high-energy peak is produced from the $\pi^0$ decay and peaks at
\eqb
x_{\rm \pi}=\eta_{\pi^0}\gamma_{\rm p,max}
\label{eq:A2}
\eqe
where $\eta_{\pi^0}\simeq 350$ \citep{DMPM12}. Finally, the medium-energy peak is produced from the synchrotron radiation of the electron-positron  secondaries from (i) $\pi^\pm$ decay and from (ii) $\gamma\gamma$ interactions between the hard photons of the $\pi^0$ peak with the soft ones from the proton synchrotron peak. It turns out that the synchrotron spectra of these two components peak at very similar energies. Taking, for simplicity, the second one we find
\eqb
x_{\rm \pm,syn}\simeq b \left( \frac{x_{\rm \pi}}{2} \right )^{2}\simeq \frac{1}{4}b\eta_{\pi^0}^2\gamma_{\rm p,max}^2,
\label{eq:A3}
\eqe 
which, interestingly enough, lies in the GeV-TeV band for a wide range of parameters. 

A necessary condition for the above picture to occur is that proton synchrotron photons are energetic enough as to pion-produce on the protons. This gives a rather strong lower limit on $\gamma_{\rm p,max}$ 
\eqb
\gamma_{\rm p,max}\ge \left({\frac{m_{\rm \pi} m_{\rm p}}{m_{\rm e}^2}\left (1+\frac{m_{\rm \pi}}{2m_{\rm p}}\right )}\right)^{1/3}b^{-1/3}
\label{eq:A4}
\eqe
or, numerically,
\eqb
\gamma_{\rm p,max}=2.8\times10^6B^{-1/3}
\label{eq:A5}
\eqe
where $B$ is in Gauss. Note that when the above condition is met, then the $\gamma$-rays from $\pi^0$ decay will also be above the threshold for pair-production on the proton synchrotron photons; therefore, the middle peak will be supplied by both channels (i) and (ii). 

At early times, well before a steady state is established  (lower curves), the luminosity of the second peak is low, but grows approximately quadratically with time, as it is supplied by an increasing rate of secondaries. The fact that the steady-state spectrum, computed for a subset of processes (electron and proton synchrotron radiation, photopion production, and $\gamma\gamma$ pair production), is very similar to the steady state solution obtained when all processes are taken into account, suggests that only a few process determine the photon emission. For example, the position of the peaks in the MW spectrum appear identical and it is only the Bethe-Heitler production that produces extra radiation that fills partly the region between the proton synchrotron and secondary pairs peak.  

Finally, the shaded region in Fig.~\ref{Fig:A1} (see equation 34 of \citealt{PM11} or equation \ref{eq9} in this paper) indicates a ``forbidden region'' for the photon emission, in the sense that $\gamma-$rays cannot exist there in steady state. Whenever the compactness of $\gamma-$ray photons falls within this region, $\gamma-$rays are automatically quenched \citep{Stawarz07,PM11}, i.e., the $\gamma$-rays annihilate on the synchrotron radiation of spontaneously created electron-positron pairs. For the parameters used in this example, the steady state spectrum lies well below this region. In the next section, we discuss the photon emission in the supercritical regime, by increasing $\ell_{\rm p}$ and forcing the $\gamma-$ray photons to enter this forbidden area. 
\begin{figure*}
 \centering
 \includegraphics[width=\linewidth]{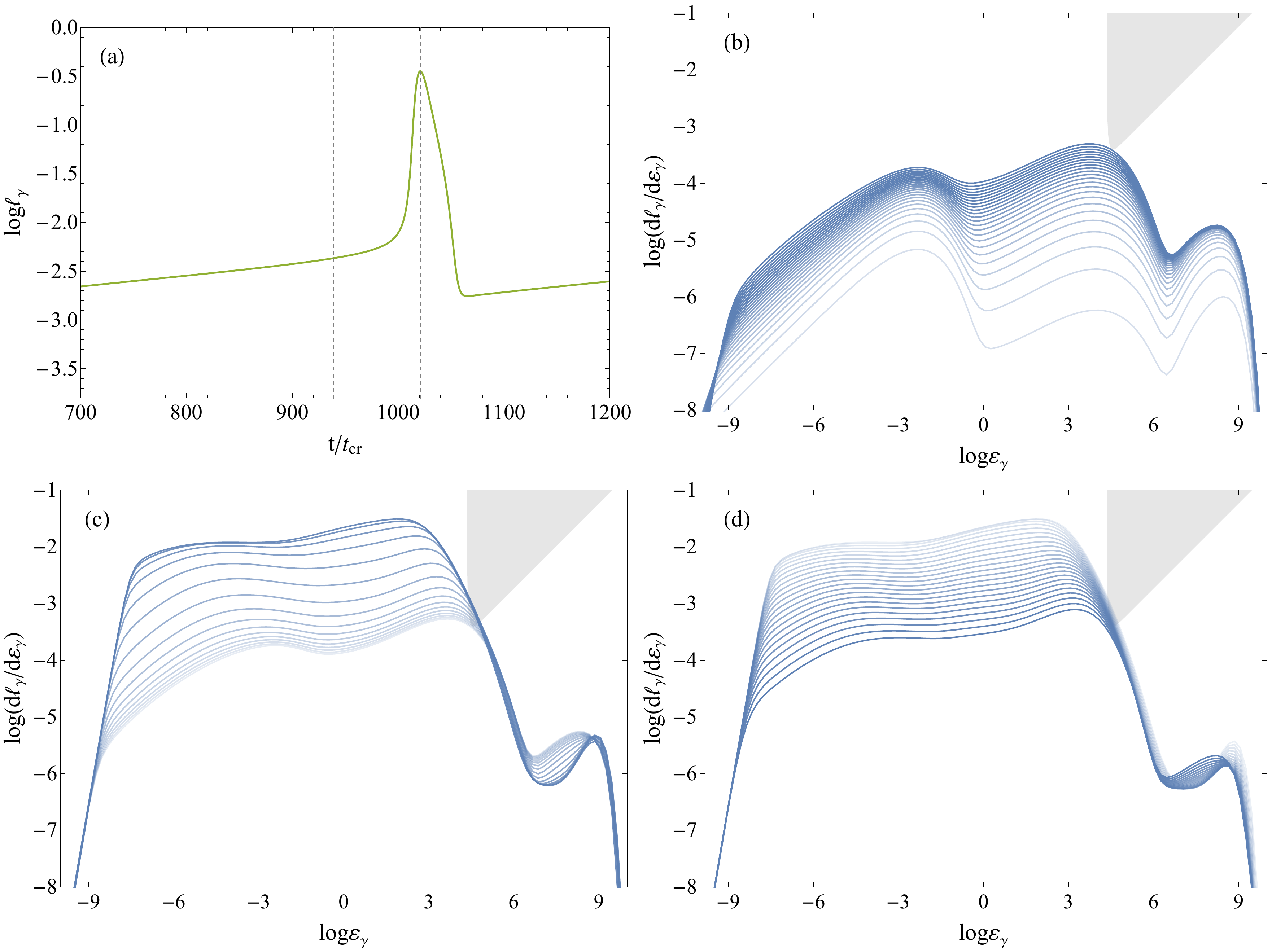}
 \caption{Zoom-in at the first peak of the supercritical light curve with $\log\ell_{\rm p}=-4.125$ shown in Fig.~\ref{Fig:1}. The times marked with vertical dashed lines denote the areas where the multi-wavelength spectrum approaches the quenching regime (panel b), the flaring state (panel c), and the decay phase (panel d). The various curves in panels (b) to (d) show snapshots from the following time intervals: $(0-939) \tcr$ (panel b), $(939-1021) \tcr$  (panel c), and $(1021-1070) \tcr$ (panel d). Time intervals between snapshots are not the same between panels.}
 \label{Fig:A2}
\end{figure*}
\subsection{Supercritical photon emission and the $\gamma Q$ loop}
Fig.~\ref{Fig:A2} shows a zoom-in to the top light curve depicted in Fig.~\ref{Fig:1} around the first peak for $\ell_{\rm p}=10^{-4.125}$ (panel a). It also plots the MW spectra before the flare (panel b), at the peak of the flare (panel c), and at the decay phase of the flare (panel d). The spectra shown in panel (a) have not yet reached the quenching regime (shaded region) and are  similar to those shown in Fig.~\ref{Fig:A1}.
However, once the $\gamma-$rays enter this regime, the $\gamma Q$ loop (first described in PM12) starts operating. The excess $\gamma-$rays turn spontaneously into electron-positron pairs which emit synchrotron photons. These synchrotron photons, on the one hand, contribute to the attenuation of $\gamma-$rays but, more importantly, cool the protons via photopion production, and as a consequence more pions are produced, which in turn results in the production of more secondary $\gamma-$rays. Therefore, a strong feedback loop, which starts as soon as the $\gamma-$rays enter the quenching regime, is developed that explosively removes part of the energy stored in protons. 
In panel (c), the MW spectrum is dominated by the leptonic secondaries and also by strong $\gamma\gamma$ attenuation due to the increasing photon compactness. Therefore, at the peak of the flare, the photon spectrum peaks at rather low $\gamma-$ray energies. The photon outburst continues either until the high-energy protons cool or the $\gamma-$ray emission moves out of the quenching zone. Once one of the above occurs, the feedback stops operating and the non-linearly produced photons escape freely from the system (panel d). However, since we have assumed a constant injection of protons, the cycle will start all over again producing, eventually, a limit-cycle behaviour. Note in panel (d) that, at the end of the cycle, the photon compactness has dropped below the quenching regime.

Finally, one can approximately derive the necessary conditions for the loop described above to operate. The first condition requires that the energy of photons in the pair synchrotron peak is above the threshold for automatic $\gamma-$quenching which is \citep{Stawarz07, PM11}
\eqb
x_{\rm \pm,syn}\ge \left ( \frac{8}{b}\right )^{1/3}.
\label{eq:A6}
\eqe
Using equation \ref{eq:A3} together with the previous relation one finds a lower limit for the proton maximum Lorentz factor
\eqb
\gamma_{\rm p,max}\ge \frac{2^{3/2}}{\eta_{\rm \pi^0}}b^{-2/3}
\label{eq:A7}
\eqe
or numerically
\eqb
\gamma_{\rm p,max}\ge 5.6\times10^6B^{-2/3}.
\label{eq:A8}
\eqe

The second condition is that the non-linear photons which are created by the spontaneous quenching of the $\gamma-$rays (equation \ref{eq:A3}) are energetic enough as to pion produce on the protons. Given that the non-linear photons are created at energies 
\eqb
x_{\rm q}=b\gamma^2_{\rm q}=\frac{b}{4} x_{\pm,syn}^{2},
\label{eq:A9}
\eqe
this condition yields
\eqb
\gamma_{\rm p,max}\ge
\left( \frac{64}{\eta_{\rm \pi^0}^{4}} \frac{m_{\rm \pi}}{m_{\rm e}}\left (1+\frac{m_{\rm \pi}}{2m_{\rm p}}\right ) \right)^{1/5}b^{-3/5}
\label{eq:A10}
\eqe
or numerically
\eqb
\gamma_{\rm p,max}\ge 9.6\times10^6B^{-3/5}.
\label{eq:A11}
\eqe
Therefore, for $B=10^{1.5}$ G (see also Fig.~\ref{Fig:5}), we expect that the $\gamma Q$ loop should start operating at $\gamma_{\rm p,max}\simeq 10^6$, which is indeed the case. Eqs.~\ref{eq:A8} and \ref{eq:A11} are only necessary, but not sufficient, conditions for driving the system into supercriticality. In addition to the conditions \ref{eq:A8} and \ref{eq:A11}, the proton density or equivalently the proton injection compactness needs to exceed a certain value for the system to become supercritical (sufficient condition).


\label{lastpage}
\end{document}